\documentclass[aps, prd,twocolumn,superscriptaddress,nofootinbib,preprintnumbers]{revtex4-1}
\def\apj{ApJ}%
\def\apjl{ApJ}%
\def\apjs{ApJS}%
\def\aap{A\&A}%
\def\mnras{MNRAS}%
\def\prd{Phys.~Rev.~D}%
\def\jcap{JCAP}%
\def\physrep{Phys.~Rep.}%
\def\procspie{Proc.~SPIE}%
\include{notations}
\usepackage{slashed}
\usepackage{graphicx}
\graphicspath{{./figure/}}
\usepackage{grffile}
\usepackage[normalem]{ulem}
\usepackage{dcolumn}
\usepackage{xcolor}
\usepackage{booktabs}
\usepackage{amsmath}
\usepackage{amssymb}
\usepackage{slashed}
\usepackage{todonotes}
\usepackage{comment}
\usepackage{float}
\usepackage{natbib}
\usepackage{hyperref}
\usepackage{mathtools}

\usepackage{tabularx}

\usepackage{comment}

\newcommand{\beq}{\begin{equation}}
\newcommand{\eeq}{\end{equation}}
\newcommand{\beqa}{\begin{eqnarray}}
\newcommand{\eeqa}{\end{eqnarray}}

\usepackage[T1]{fontenc}
\usepackage{ae,aecompl}
\usepackage{lineno}
\usepackage{eso-pic}% http://ctan.org/pkg/eso-pic
\usepackage{txfonts}

\defcitealias{Battaglia:2012}{B12}
\defcitealias{Vikram2017}{V17}

\usepackage{lineno}
\begin{document}

\title[Future halo-$y$ correlations]{Constraining the properties of gaseous halos via cross-correlations of upcoming galaxy surveys and thermal Sunyaev-Zel'dovich maps}

\author{S.~Pandey}
\affiliation{Department of Physics and Astronomy, University of Pennsylvania, Philadelphia, PA 19104, USA}
\author{E.~J.~Baxter}
\affiliation{Institute for Astronomy, University of Hawai'i, 2680 Woodlawn Drive, Honolulu, HI 96822, USA}
\affiliation{Kavli Institute for Cosmology Cambridge, Institute of Astronomy, University of Cambridge, Cambridge CB3 0HA, UK}
\affiliation{Department of Physics and Astronomy, University of Pennsylvania, Philadelphia, PA 19104, USA}
\author{J.~C.~Hill}
\affiliation{Department of Physics, Columbia University, New York, NY 10027, USA}
\affiliation{Flatiron Institute, Center for Computational Astrophysics, New York, NY 10010, USA}
\affiliation{School of Natural Sciences, Institute for Advanced Study, Princeton, NJ 08540, USA}

% \date{Last updated \today}

\label{firstpage}

\begin{abstract}
The thermal Sunyaev-Zel'dovich (tSZ) effect induces a Compton-$y$ distortion in cosmic microwave background (CMB) temperature maps that is sensitive to a line of sight integral of the ionized gas pressure.  By correlating the positions of galaxies with maps of the Compton-$y$ distortion, one can probe baryonic feedback processes and study the thermodynamic properties of a significant fraction of the gas in the Universe.  Using a model fitting approach, we forecast how well future galaxy and CMB surveys will be able to measure these correlations, and show that powerful constraints on halo pressure profiles can be obtained.  Our forecasts are focused on correlations between galaxies and halos identified by the upcoming Dark Energy Spectroscopic Instrument survey and tSZ maps from the Simons Observatory and CMB-S4 experiments, but have general applicability to other surveys, such as the Large Synoptic Survey Telescope.  We include prescriptions for observational systematics, such as halo miscentering and halo mass bias, demonstrating several important degeneracies with pressure profile parameters.  Assuming modest priors on these systematics, we find that measurements of halo-$y$ and galaxy-$y$ correlations with future surveys will yield tight constraints on the pressure profiles of group-scale dark matter halos, and enable current feedback models to either be confirmed or ruled out.  
\end{abstract}

\maketitle

\section{Introduction}
\label{sec:intro}

The thermal Sunyaev-Zel'dovich (tSZ) effect results from inverse Compton scattering of cosmic microwave background (CMB) photons with hot ionized gas \citep{SZ_1972}.  This process leads to a spectral distortion in the CMB that is measured by current CMB experiments.  The amplitude of the tSZ distortion in the CMB is proportional to the Compton-$y$ parameter, which is sensitive to the line-of-sight integral of the ionized gas pressure. Since massive galaxy clusters are reservoirs of hot gas, the tSZ effect has long been used to detect galaxy clusters, making it a powerful cosmological probe through its sensitivity to dark matter halo abundance \citep[e.g.][]{Carlstrom:2002}.  This sensitivity to halo abundance also leads to the angular power spectrum of $y$ being an especially sensitive probe of structure, with its amplitude scaling roughly as $\sigma_8^8$ \citep[e.g.][]{Komatsu:2002,Hill-Pajer2013,Horowitz-Seljak2017,Bolliet2019}.

In addition to {\it detecting} halos, the tSZ effect is also a useful tool for studying the pressure profiles of the baryons {\it within} the halos.  Because $y$ is a line-of-sight integrated quantity, however, it is difficult to use measurements of $y$ alone to study the halo mass or redshift dependence of these profiles.  Furthermore, the angular power spectrum of $y$ receives a dominant contribution from the most massive halos, making it particularly difficult to probe the pressure profiles of lower-mass halos with this observable.  Cross-correlations of $y$ with tracers of the density field, on the other hand, can be used to isolate contributions to $y$ from halos of a particular mass or redshift range~\citep[e.g.][]{Planck:LBG,Greco:2015,Hill-Spergel2014,VanWaerbeke:2014,Battaglia2015,Hojjati2015,Vikram2017,Hojjati2017,Makiya:2018,Hill:2018}. 

Constraining the distribution and energetics of the baryons within and around halos is important for improving our understanding of astrophysical feedback, which is responsible for preventing the over-formation of stars in galaxies \citep[for a review see, e.g.][]{Benson:2010}.  Various feedback mechanisms have been proposed, such as feedback from active galactic nuclei \citep[AGN; ][]{Benson:2003}, believed to be important for high-mass halos, and supernovae and stellar winds \citep[e.g.][]{Efstathiou:2000, Fierlinger:2016}, believed to be most important for low-mass halos.  In general, these mechanisms inject energy and momentum into the halo gas, thereby preventing the gas from cooling to form stars; however, the details of these mechanisms and their relative importance as a function of halo mass and redshift are not well constrained.  

Because feedback can both inject energy into the gas and change its distribution, one would expect feedback to change the pressure profile of the gas, and thereby lead to an observational signature via the tSZ effect \citep{Battaglia:2017,Battaglia:2019}. Feedback processes are expected to have larger impact on lower-mass halos, since for these objects, the fractional contribution of feedback energy to the total thermodynamic energy budget is expected to be larger, allowing for more dramatic effects such as expulsion of gas from the halo~\citep[e.g.][]{VandeVoort2016}.  One of the main goals of this work is to make forecasts for how well future measurements of  halo-$y$ correlations can constrain the amplitude and shape of the halo pressure profiles for group-scale dark matter halos.  

In addition to its importance for galaxy formation, feedback also impacts the matter power spectrum on small scales (roughly $1$--$10$\% changes for wavenumbers $0.5 \, h/$Mpc $\lesssim k \lesssim 5 \, h/$Mpc~\citep[e.g.][]{vanDaalen:2011}), making it an important systematic for cosmological constraints from e.g. weak lensing surveys \citep{Eifler:2015,Chisari2019}.  Constraints on feedback models from halo-$y$ correlations should enable the signal-to-noise of weak lensing surveys at small scales to be better exploited, thereby enhancing cosmological constraints.

To date, several studies have used measurements of the tSZ signal to constrain astrophysical feedback.  An early analysis from {\it Planck} measured the stacked tSZ signal of ``locally brightest galaxies'' extracted from SDSS data, finding (surprisingly) that the integrated Compton-$y$ -- halo mass relation was consistent with self-similar theoretical predictions~\citep{Planck:LBG}.  Subsequent analyses showed that this was a result of the coarse angular resolution of {\it Planck}, and that the results were also consistent with models incorporating AGN feedback~\citep{LeBrun15,Greco:2015}.  \citet{Hill:2018} pointed out that these results could also be biased due to their neglect of the two-halo term in the theoretical interpretation.  Amongst several other studies in recent years, \citet{Soergel2017} placed upper limits on AGN feedback using measurements of the tSZ signal near BOSS-identified quasars using {\it Planck} data (see also~\citep{Verdier2016}).  \citet{Pandey:2019} used high signal-to-noise cross-correlations of galaxies detected by the Dark Energy Survey with {\it Planck} $y$-maps to constrain energy injection  from feedback as a function of redshift.  Recently,  \citet{Hall2019} used measurements of the  tSZ signal around quasars at $z \sim 2$ with the Atacama Cosmology Telescope (ACT) to constrain quasar feedback.

We take a model fitting approach in our forecasts, in which we adopt parameterized forms for the halo pressure profiles and their dependence on halo properties, and generate forecasts for how well these parameters can be constrained.  An alternative to this approach would be to generate forecasts only for the halo-$y$ correlations themselves, without forecasting constraints on model parameters.  The advantages of the adopted approach are that it allows us to consider joint constraints with multiple observables, and that it can easily account for the effects of important systematics, like halo miscentering and mass bias.  Moreover, the model fitting approach allows the 3D pressure profiles of dark matter halos to be inferred from the inherently 2D observations of Compton-$y$.  Finally, this approach reflects how we expect future analyses with data to actually proceed. 

Measurements of halo-$y$ correlations are expected to improve significantly in the next decade.  Upcoming CMB surveys like Simons Observatory \citep[SO;][]{Ade:2019} and CMB Stage 4 \citep[CMB-S4; ][]{Abazajian:2016} will map the sky at submillimeter frequencies to significantly greater depth and at significantly higher resolution than {\it Planck} \citep{PlanckOverview}.  At the same time, upcoming surveys like the Dark Energy Spectroscopic Instrument \citep[DESI; ][]{DESI:FDR} and the Large Synoptic Survey Telescope \citep[LSST; ][]{LSST:sciencebook} will yield galaxy catalogs with larger area and out to higher redshift than previous surveys.  With improvements on both the CMB and galaxy survey side, we forecast detection significances for some halo-$y$ correlation measurements in excess of $500\sigma$.  Additionally, as we show below, relative to current $y$ maps with {\it Planck}~\cite{Planck:tsz}, future measurements will have improved angular resolution, enabling the baryon distribution in low-mass halos to be probed. 

Of course, improvements in the signal-to-noise of halo-$y$ correlations will be meaningless if sources of systematic errors cannot be controlled.  In the modeling framework that we use in this analysis, we parameterize important sources of systematic error, namely halo miscentering and bias in the mass constraints for the populations of halos.  We show that these systematics can significantly degrade pressure profile constraints.  However, with fairly modest assumptions about the level at which these systematics can be controlled, tight constraints on halo pressure profiles can be achieved.  Furthermore, for the case of the most massive halos, the halo-$y$ correlation measurements themselves are sufficient to self-calibrate some of these systematics. We also comment on systematics impacting the $y$-maps from future surveys, such as contamination from the cosmic infrared background.
 
Given a galaxy catalog, one could imagine pursuing two different strategies to measure the halo-$y$ correlation.  In the first, one identifies bound dark matter halos (for instance, by applying a friends-of-friends algorithm to the galaxy catalog), and then correlates the resultant halo sample with a Compton-$y$ map to directly constrain the halo pressure profile.  This is the approach taken by, e.g. \citet{Vikram2017} and \citet{Hill:2018}. An alternative approach is to directly correlate a galaxy catalog with a Compton-$y$ map, and use a parameterized relationship between the galaxies and halos (i.e., a halo occupation distribution, or HOD) to constrain the halo pressure profiles.  This is the approach taken by, e.g. \citet{Makiya:2018}. We will present forecasts for both types of analyses below.  For the galaxy-based forecasts, an important question is how degenerate the HOD model is with the parameters characterizing the pressure profiles.  Note that \citet{Pandey:2019} also correlated galaxies with Compton-$y$ maps to constrain feedback physics, but because the fitting was restricted to the two-halo regime, precision modeling of the galaxy-halo connection was not necessary.

The paper is organized as follows.  In \S\ref{sec:formalism}, we present the formalism for modeling the relevant auto- and cross-power spectra between the halo field and the Compton-$y$ field.  Our main results are presented in two sections: in \S\ref{sec:halo_results}, we present forecasts for the halo-based approach; in \S\ref{sec:galaxy_results}, we present forecasts for the galaxy-based approach.  We conclude in \S\ref{sec:discussion}.  When adopting a fiducial cosmological model, we will assume flat $\Lambda$CDM  with $h = 0.677$, $\Omega_{\rm m} = 0.315$, $\Omega_b=0.0486$, $n_s = 0.9667$ and $\sigma_8 = 0.816$, matching the best-fit parameters from \citet{Planck:2018cosmo}.

\section{Forecasting methodology}
\label{sec:formalism}

We are interested in how cross-correlations between galaxies and maps of the Compton-$y$ parameter can be used to constrain models of astrophysical feedback.  To this end, we consider two different types of forecasts.  In the first, we imagine that a galaxy survey has been used to identify dark matter halos in some mass and redshift range.  In the second approach, we imagine that the galaxies themselves are directly correlated with the Compton-$y$ maps, and an HOD framework is used to relate the galaxies to the underlying halo population.  The formalism that we introduce in this section is sufficiently general to capture both types of forecasts.

We model three possible observables: the $y$-$y$, galaxy-$y$ and galaxy-galaxy correlation functions. We use the halo model approach~\citep[for a review see][]{Cooray:2002} to model these observables and their covariance, following closely the work of \citet{Komatsu:1999}, \citet{Seljak:2000}, and \citet{Makiya:2018}.

We model the harmonic-space correlations between any two probes ${\rm A}$ and ${\rm B}$ (here we consider galaxies, $g$, halos, $h$, and Compton-$y$, $y$) as the sum of a one-halo term and a two-halo term.  In the halo model framework, the one-halo term is given by an integral over redshift ($z$) and halo mass ($M$):
\begin{multline}\label{eq:Cl1h}
C_{\ell}^{\rm A,B; 1h} = \int_{z_{\rm{min}}}^{z_{\rm{max}}} dz \frac{dV}{dz d\Omega} \int_{M_{\rm{min}}}^{M_{\rm{max}}} dM \frac{dn}{dM} \bar{u}_{\ell}^{\rm A}(M,z) \ \bar{u}_{\ell}^{\rm B}(M,z),
\end{multline}
where $dV$ is the cosmological volume element, $dn/dM$ is the halo mass function, and $\bar{u}^{\rm A}_\ell$ and $\bar{u}^{\rm B}_\ell$ are the multipole-space profiles of observables A and B.  We use the \citet{Tinker:2008} fitting function for $dn/dM$ throughout.   The two-halo term is given by
\beqa\label{eq:Cl2h}
C_{\ell}^{\rm A,B; 2h} = \int_{z_{\rm{min}}}^{z_{\rm{max}}} dz \frac{dV}{dz d\Omega} b_{\ell}^{\rm A}(z) \ b_{\ell}^{\rm B}(z) \ P_{\rm{lin}}((\ell + 1/2)/\chi,z),
\eeqa
where $b^{\rm A}_\ell$ and $b^{\rm B}_\ell$ are effective linear bias parameters describing the clustering of tracers ${\rm A}$ and ${\rm B}$ respectively, $P_{\rm lin}(k,z)$ is the linear matter power spectrum and $\chi$ is the comoving distance corresponding to the redshift $z$. The exact form of $b_{\ell}$, $\bar{u}_{\ell}$ and the values of $M_{\rm{min}}$, $M_{\rm{max}}$ will depend on the particular fields being correlated; we will describe these quantities in the following sections. 

The halo model framework adopted here ignores the higher order effects like quasi-linear effects in the one-to-two halo transition regime, halo exclusion effects for galaxies and for the hot gas. We defer a more careful exploration of these issues to future work. However, we note that the one-halo term carries the majority of the constraining power for our parameters of interest, so we do not expect quasi-linear scales to impact our inferences significantly.

\subsection{Pressure profile model}\label{sec:pressure}

The Compton-$y$ parameter in some direction on the sky is given by:
\beqa
y = \frac{\sigma_T}{m_e c^2} \int_0^{\infty} dl \, P_e(l),
\eeqa 
where $P_e(l)$ is the electron gas pressure (which dominates the inverse Compton scattering process that gives rise to the SZ effect) over the line of sight distance $l$, $\sigma_T$ is the Thomson cross section, $m_e$ is the electron mass, and $c$ is the speed of light.   For a fully ionized gas consisting of hydrogen and helium, the electron pressure, $P_e$, is related to the total thermal pressure, $P_{\rm th}$, by:
\beqa \label{eq:batt1}
P_e = \left[\frac{4-2Y}{8-5Y}\right] P_{\rm th},
\eeqa
where $Y$ is the primordial helium mass fraction.

For a halo of mass $M_{200c}$ and radius $R_{200c}$ at redshift $z$, we label its pressure profile as $P_e(x|M,z)$, where $x = r/R_{200c}$ and $r$ is the radial distance.  The Fourier transformed Compton-$y$ profile is then
\begin{multline}\label{eq:uyl}
\bar{u}^{y}_{\ell}(M_{200c},z) = \frac{4\pi r_{200c}}{l^2_{200c}} \frac{\sigma_T}{m_e c^2} \int_{x_{\rm{min}}}^{x_{\rm{max}}} dx \ x^2 \ P_e(x|M_{200c},z) \\ \times \frac{\sin(\ell x/l_{200c})}{\ell x/l_{200c}},
\end{multline}
where $l_{200c} = D_A/R_{200c}$ and $D_A$ is the angular diameter distance to redshift $z$. We choose $x_{\rm{min}} = 10^{-3}$ and $x_{\rm{max}} = 5$, which ensures that the above integral captures the contribution to the pressure from the extended profile of hot gas. 

The effective tSZ bias $b_\ell^y$ is given by:
\beqa\label{eq:by}\label{eq:byl}
b_\ell^y = \int_{M_{\rm{min}}}^{M_{\rm{max}}} dM \ \frac{dn}{dM} \bar{u}_\ell^y(M,z) b_{\rm{lin}}(M,z),
\eeqa
where $b_{\rm{lin}}$ is the linear bias of halos with mass $M$ at redshift $z$.  We use the \citet{Tinker:2010} fitting function for halo bias as a function of mass and redshift.  

As a starting point for modeling the halo pressure profile, $P_e(x|M,z)$, we take the simulation-derived AGN-200c model from \citet{Battaglia:2012} (hereafter \citetalias{Battaglia:2012}), which we refer to as $P_e^B(x|M,z)$.  This fitting function is calibrated at high mass ($M_{200c} \gtrsim 5 \times 10^{13} \, M_{\odot}/h$), and may therefore be incapable of capturing the impact of feedback on lower-mass halos.  We will therefore introduce additional freedom to this model below.

The \citetalias{Battaglia:2012} pressure profile is parametrized by a generalized NFW form:
\beq\label{eq:B12_model}
P^{\rm B}_e(x|M,z) = P_{\Delta} P_0 \bigg( \frac{x}{x_c} \bigg)^{\gamma} \big[ 1 + (x/x_c)^{\lambda} \big]^{-\beta},  
\eeq
where
\begin{equation}
P_{\Delta} = \frac{G \Delta M_{\Delta} \rho_{c}(z) \Omega_b}{2 R_{\Delta} \Omega_m},
\end{equation}
for any spherical overdensity, $\Delta$, relative to the critical density, $\rho_{c}$.  We will typically use $\Delta=200$. Following \citetalias{Battaglia:2012}, we fix $\lambda = 1.0$ and $\gamma = -0.3$.  For the parameters $P_0$, $x_c$ and $\beta$, \citetalias{Battaglia:2012} adopts a scaling relation of the form:
\beq\label{eq:B12_scaling}
A(M_{200},z) = A_{\rm high} \bigg(\frac{M_{200}}{M_{{\rm high}}}\bigg)^{\alpha} (1 + z)^{\omega},
\eeq
where $A(M,z)$ generically represents some parameter, $A_{\rm high}$ is the parameter value at $M_{200} = M_{\rm high}$ and $z=0$, and $\alpha$ and $\omega$ describe the scaling of the parameter with mass and redshift, respectively.  The best-fit values of these parameters are given in Table~1 of \citetalias{Battaglia:2012}, and unless specified otherwise, we will adopt these best-fit values.  

Motivated by the results from the hydrodynamical simulations of \citet{LeBrun15}, we fix $M_{{\rm high}} = 3 \times 10^{14} M_{\odot}/h$. We note that this choice of pivot mass is different from the one used in \citetalias{Battaglia:2012}, but we have rescaled the amplitude of the pressure profile accordingly. 

For the halo-based forecasts, we will treat the amplitudes, $A_{\rm high}$, of the parameters $P_0$ and $\beta$ as free parameters.  In a slight abuse of notation, we will refer to these amplitudes as $P_0$ and $\beta$, respectively; one should remember, though, that the mass and redshift scalings of these parameters are preserved via Eq.~\ref{eq:B12_scaling}.  For the halo-based forecasts, we allow these parameters to vary separately for each mass and redshift bin.  This gives the model a large degree of freedom to capture possible departures from the \citetalias{Battaglia:2012} model as a result of, e.g. feedback.  We summarize the model parameters and priors for the halo-based forecasts in Table~\ref{tab:hy}.

In contrast to the halo-based forecasts, the galaxy-$y$ cross-correlations receive contributions from a very wide range of halo masses.  When fitting these correlation functions, making the model choice described above is too restrictive: keeping only $P_0$ and $\beta$ free in fits to the galaxy-$y$ correlations would essentially not allow for any feedback effects at low halo mass.  When fitting the galaxy-$y$ correlations, we therefore consider a modified version of the \citetalias{Battaglia:2012} profile:
\beqa\label{eq:Pe_total}
P_e(r|M,z) \rightarrow
\begin{cases}
P^{\rm B}_e(r|M,z) \,, \ \ \ M \geq M_{\rm high} \\
P^{\rm B}_e(r|M,z) \left( \frac{M}{M_{\rm high}} \right)^{\alpha_{p, {\rm mid} }},  M_{\rm low} < M < M_{\rm high} \\
P^{\rm B}_e(r|M,z) \left( \frac{M}{M_{\rm high}} \right)^{\alpha_{ p, {\rm mid} }} \, \left( \frac{M}{M_{\rm low}} \right)^{\alpha_{p, {\rm low} }},  M < M_{\rm low}
\, \nonumber
\end{cases}
\\
\label{eq:UB}
\eeqa
where we choose $M_{\rm low} = 3 \times 10^{13} M_{\odot}/h$.  In this model, we will refer to the $\alpha$ (Eq.~\ref{eq:B12_scaling}) parameter for $P_0$ as $\alpha_{p,{\rm high}}$.  By allowing $\alpha_{p,{\rm mid}}$ and $\alpha_{p,{\rm low}}$ to be free, we allow significant freedom in the \citetalias{Battaglia:2012} model at low halo masses.

For the galaxy-based forecasts, we will treat $P_0$ and $\beta$ as free parameters, as well as the three mass scaling parameters $\alpha_{ p, {\rm high} }$, $\alpha_{ p, {\rm mid} }$ and $\alpha_{ p, {\rm low} }$. The fiducial values and description of priors of these parameters are given in Table \ref{tab:gy}. 

\subsection{Models for the galaxy and halo distributions}\label{sec:hod_halo}

We now model the distribution of galaxies within the dark matter halos.  The Fourier transformed galaxy profile for any halo having virial mass $M_{\rm vir}$ is given by \citep{Seljak:2000}:
\begin{multline}\label{eq:ugl}
    \bar{u}^{g}_{\ell}(M_{\rm vir},z) = \frac{W^g(z)}{\chi^2} \frac{1}{\langle n_g(z) \rangle} \times \\  \sqrt{ 2 f_{\rm{cen}}(M_{\rm{vir}})  N_{\rm{sat}}(M_{\rm{vir}}) u_{\rm{sat}}(k,M_{\rm{vir}}) +  f_{\rm{cen}} N_{\rm{sat}}^2(M_{\rm{vir}}) u_{\rm{sat}}^2(k,M_{\rm{vir}}) }, 
\end{multline}
where $k = (\ell + 1/2) / \chi$, $\chi$ is the comoving distance to redshift z, $W^g = (dn_g/dz)(dz/d\chi)$ with $(dn_g/dz)$ the normalized redshift distribution of the galaxies, $f_{\rm cen}$ is the central fraction, and $N_{\rm sat}$ is the satellite occupation number.  We assume that the spatial distribution of satellite galaxies, $u_{\rm sat}$, can be approximated by a Navarro-Frenk-White (NFW) profile \citep{nfw:1996}:
\beqa
u_{\rm{sat}}(r,M_{\rm{vir}}) \propto \frac{1}{(r/r_{s,g})(r/r_{s,g} + 1)^2} \Theta (r_{\rm{max},g} - r),
\eeqa
where $\Theta$ is a Heaviside function. The scale radius of the galaxy distribution, $r_{s,g}$, is approximated to be proportional to the scale radius of matter, $r_{s}$, with $r_{s,g} = 1.17 r_{s}$ \citep{Makiya:2018}.  We relate $r_s$ to the virial radius of the halo, $r_{\rm{vir}}$, using the halo-concentration relation of \citet{Diemer19}. Also, we assume that the maximum  radius of the galaxy distribution is $r_{\rm{max},g} = r_{\rm{vir}}$. To estimate the viral mass for the halos of mass $M_{200c}$, we use the virial overdensity definition varying with redshift as described in \citet{Bryan_Norman:1998}. 

We adopt the following forms for the central fraction and satellite occuption number:
\beqa\label{eq:fcen_hod}
f_{\rm{cen}}(M_{\rm{vir}}) = \frac{1}{2} \bigg[ 1 + \rm{erf}\bigg( \frac{\log M_{\rm{vir}} - \log M_{\rm{min}}}{\sigma_{\log M_{\rm{vir}}}} \bigg) \bigg]
\eeqa
and 
\beqa\label{eq:nsat_hod}
N_{\rm{sat}}(M_{\rm{vir}}) = \bigg( \frac{M_{\rm{vir}} - M_0}{M_1} \bigg)^{\alpha_g} \Theta (M_{\rm{vir}} - M_0).
\eeqa
These forms are motivated by the SDSS analyses of \citet{Zheng:HOD} and \citet{Zehavi:2011}.  We fix $\sigma_{\log M_{\rm{vir}}} = 0.17$ and $\log M_{\rm{min}} = 11.57$ as described in Table 3 of \citet{Zehavi:2011} for the volume limited galaxy sample having absolute magnitude less than -19.5, the forecasted maximum absolute magnitude of the DESI  Bright Galaxy Survey (BGS) sample. The parameter $M_0$ denotes the minimum mass a halo should have to host a satellite galaxy,  $M_1$ is the pivot mass of power law scaling relation and $\alpha_g$ is the power law index.  We will treat $M_0$, $M_1$ and $\alpha_g$ as free parameters, with fiducial values of $\log M_0 = 12.23$, $\log M_1 = 12.75$ and $\alpha_g = 0.99$, as summarized in Table~\ref{tab:gy}. 

The mean number of galaxies, $\langle n_g(z) \rangle$, entering into Eq.~\ref{eq:ugl}, is then given by:
\beqa\label{eq:ng}
\langle n_g(z) \rangle = \int_{M_{\rm{{min}}}}^{M_{\rm{{max}}}} dM \frac{dn}{dM} f_{\rm{cen}}(M_{\rm{vir}}) (1 + N_{\rm{sat}}(M_{\rm{vir}})),
\eeqa
where, $M_{\rm{{min}}}$ and  $M_{\rm{{max}}}$ corresponds to the boundary of a particular mass bin.  Similarly, the effective large scale bias of the galaxies is given by:
\begin{multline}\label{eq:bgl}
    b^{g}_{\ell}(z) = \frac{W^g(z)}{\chi^2} \frac{1}{\langle n_g(z) \rangle} \int_{M_{\rm{{min}}}}^{M_{\rm{{max}}}} dM \frac{dn}{dM} \\
    f_{\rm{cen}}(M_{\rm{vir}}) (1 + N_{\rm{sat}}(M_{\rm{vir}})u_{\rm{sat}}(k,M_{\rm{vir}}))b_{\rm{lin}}(M,z).
\end{multline}

While we do not include halo clustering as one of the data vectors used in our analysis, we must model this quantity in order to compute the covariance of the halo-$y$ correlation (see next section).  Halos are, by definition, treated as point objects when measuring their auto-correlation. Therefore, for the halo-halo correlation, all the contribution comes from the two-halo term. The effective bias of the halos is:
\beqa\label{eq:bhl}
    b^h_{\ell}(z) = \frac{W^h(z)}{\chi^2} \frac{1}{\langle n_h(z) \rangle} \int_{M_{\rm{{min}}}}^{M_{\rm{{max}}}} dM \frac{dn}{dM} b_{\rm{lin}}(M,z),
\eeqa
where
\begin{eqnarray}
n_h(z) &=& \int_{M_{\rm{{min}}}}^{M_{\rm{{max}}}} dM \frac{dn}{dM} \\
W^h &=& \frac{dn_h}{dz}\frac{dz}{d\chi}
\end{eqnarray}
where $(dn_h/dz)$ is the normalized redshift distribution of the halos.

\begin{table}
  \caption{Parameters varied in the halo-based forecasts, along with their fiducial values and priors ($\mathcal{U} \equiv$ uniform prior and $\mathcal{N} \equiv$ Gaussian prior).}
  \begin{tabularx}{0.45\textwidth}{llc}
    \hline\hline
    Parameter & Value and Prior  & Meaning \\
    \hline
    $P_0$ & 18.1, $\mathcal{U}[0,80]$  & Pressure profile Amplitude (Eq.~\ref{eq:B12_model}) \\
    $\beta$ & 4.35, $\mathcal{U}[1,8]$ & Shape of pressure profile (Eq.~\ref{eq:B12_model})  \\
    $\ln c_{\rm mis}$ & -1.1, $\mathcal{N}[\sigma=0.2]$ & Miscentering distance (Eq.~\ref{eq:mis_sig})  \\
    $f_{\rm mis}$ & 0.2, $\mathcal{N}[\sigma=0.1]$ & Miscentering fraction (Eq.~\ref{eq:mis_f})  \\
    $\eta$ & 1.0, $\mathcal{N}[\sigma=0.1]$ & Mass bias (Eq.~\ref{eq:MB_x}) \\
    \hline\hline
    \label{tab:hy}
  \end{tabularx}
\end{table}

\begin{table*}
  \caption{Parameters varied in the galaxy-based forecasts, along with their fiducial values and priors ($\mathcal{U} \equiv$ uniform prior).}
  \begin{tabularx}{0.6\textwidth}{llc}
    \hline\hline
    Parameter & Value and Prior & Meaning \\
    \hline
    $P_0$ & 18.1, $\mathcal{U}[0,80]$ & Pressure profile amplitude (Eq.~\ref{eq:B12_scaling}) \\
    $\beta$ & 4.35, $\mathcal{U}[1,8]$ & Shape of pressure profile (Eq.~\ref{eq:B12_scaling})  \\
    $\alpha_{p, {\rm high}}$ & 0.154, $\mathcal{U}[-1,1]$  & $P_e$-$M_h$ power law index (Eq.~\ref{eq:B12_scaling}) \\
    $\alpha_{p, {\rm mid}}$ & 0.0, $\mathcal{U}[-1.5,1.5]$   & $P_e$-$M_h$ power law index (Eq.~\ref{eq:Pe_total}) \\
    $\alpha_{p, {\rm low}}$ & 0.0, $\mathcal{U}[-2,2]$   & $P_e$-$M_h$ power law index for (Eq.~\ref{eq:Pe_total}) \\
    $\log M_0$ & 12.23, $\mathcal{U}[9,16]$   & Minimum mass of halo to host a satellite (Eq.~\ref{eq:nsat_hod})    \\
    $\log M_1$ & 12.75, $\mathcal{U}[9,16]$   &  Pivot mass of $N_{\rm sat}$-$M_h$ relation (Eq.~\ref{eq:nsat_hod})   \\ 
    $\alpha_g$ & 0.99, $\mathcal{U}[0.1,1.9]$   &  Scaling index of $N_{\rm sat}$-$M_h$ relation  (Eq.~\ref{eq:nsat_hod})  \\
    \hline\hline
    \label{tab:gy}
  \end{tabularx}
\end{table*}

\subsection{Covariance}
We model the covariance, $\varmathbb{C}$, of the galaxy/halo and $y$ auto- and cross-spectra as a sum of Gaussian ($\varmathbb{C}^{\rm G}$) and non-Gaussian ($\varmathbb{C}^{\rm NG}$) terms as follows:
\beq\label{eq:cov_tot}
\varmathbb{C} (C^{\rm A,B}_{\ell_1},C^{\rm C,D}_{\ell_2}) = \varmathbb{C}^{\rm G} (C^{\rm A,B}_{\ell_1},C^{\rm C,D}_{\ell_2}) + \varmathbb{C}^{\rm NG} (C^{\rm A,B}_{\ell_1},C^{\rm C,D}_{\ell_2}),
\eeq
where $A$ and $B$ represent either the galaxy/halo or $y$ fields.

The Gaussian term is given by \citep{Hu:2004}:
\beq\label{eq:cov_g}
\varmathbb{C}^{\rm G} (C^{\rm A,B}_{\ell_1},C^{\rm C,D}_{\ell_2}) = \frac{\delta_{\ell_1 \ell_2}}{f^{\rm A,B ; C,D}_{\rm sky} (2 \ell_1 + 1)\Delta \ell_1} \bigg[ \hat{C}^{\rm A,C}_{\ell_1} \hat{C}^{\rm B,D}_{\ell_2} + \hat{C}^{\rm A,D}_{\ell_1} \hat{C}^{\rm B,C}_{\ell_2}  \bigg].
\eeq
Here, $\delta_{\ell_1 \ell_2}$ is the Kronecker delta, $f^{\rm A,B ; C,D}_{\rm sky}$ is the effective sky coverage fraction, $\Delta \ell_1$ is the size of the multipole bin, and $\hat{C}_{\ell}$ is the total cross-spectrum between any pair of fields including the noise contribution. For the auto-correlation of galaxies and halos, the noise spectrum is pure shot noise, given by $1/\bar{n}$, where $\bar{n}$ is the forecasted number density of the objects. For the auto-correlation of Compton-$y$, we use the forecasted noise power spectra described below~\cite{CMBS4DSR}. 

The non-Gaussian term is well approximated by the one-halo term of the trispectrum \citep{Cooray:2001, Komatsu:2002}:
\beq\label{eq:cov_ng}
\varmathbb{C}^{\rm NG} (C^{\rm A,B}_{\ell_1},C^{\rm C,D}_{\ell_2}) = \frac{1}{4 \pi f^{\rm A,B ; C,D}_{\rm sky}} \varmathbb{T}^{\rm A,B ; C,D}_{\ell_1 \ell_2}.
\eeq
The one-halo term of the trispectrum $\varmathbb{T}$ is given by:
\beq\label{eq:cov_trispec}
\varmathbb{T}^{\rm A,B ; C,D}_{\ell_1 \ell_2} = \int dz \frac{dV}{dz d\Omega} dM \frac{dn}{dM} \bar{u}_{\ell}^{\rm A} \bar{u}_{\ell}^{\rm B} \bar{u}_{\ell}^{\rm C} \bar{u}_{\ell}^{\rm D}. 
\eeq
As mentioned above, the one-halo term should not contribute to the trispectrum of correlations involving halos, hence we assume that the $\varmathbb{C}^{\rm NG} = 0$ for that case.

\subsection{Systematics parameterizations}\label{sec:sys_all}

We now incorporate important potential sources of systematic error into our analysis.  We focus on halo miscentering and mass bias, both of which impact the halo-based constraints.  We identify these as important systematics based on the results of e.g. \citet{Vikram2017}.  

\subsubsection{Miscentering}
\label{sec:miscentering}

Our halo-based forecasts assume that the galaxy distribution has been used to identify the locations of halos.  In this process, some assumptions must be made about the centers of these halos \citep[e.g.][]{Yang:2007}.  Frequently, the halo center is chosen to be at the location of the brightest galaxy in the identified halo.  However, this prescription may not yield the true halo center.  Any difference between the assumed halo center and the true halo center  (i.e. miscentering) will then lead to smearing of the halo-$y$ correlation at small scales (see Fig.~\ref{fig:Cells}).  For the galaxy-based forecasts, on the other hand, a prescription for miscentering is not needed, since the modeling formalism effectively parameterizes the distribution of galaxies within the halo.

Since miscentering happens in real-space, we incorporate its effect by first transforming the halo-$y$ cross-spectrum into a correlation function.  The halo-$y$ correlation at any projected distance, $R$, for the halos in a redshift and mass bin can be obtained by the Hankel transform of the harmonic space power spectrum via:
\beqa
\xi^{\rm{halo}-y}(R) = \int d\ell \ \ell \ C^{\rm{halo}-y}_{\ell} \ J_0 \bigg(\ell \ \frac{R}{\chi(\bar{z})} \bigg).
\eeqa
Here $J_0$ is the zeroth order Bessel function of first kind and $\bar{z}$ is the mean redshift of the halos in the tomographic bin. If the center of halos are incorrectly estimated by a distance of $R_{\rm mis}$ from the true center, the miscentered profile will be given by:
\begin{multline}
\xi^{\rm{halo}-y}(R|R_{\rm mis}) = \\
\int_0^{2\pi} \frac{d\theta}{2\pi} \xi^{\rm{halo}-y} \bigg(\sqrt{R^2 + R^2_{\rm mis} + 2RR_{\rm mis}\cos\theta} \bigg) .
\end{multline}
We can approximate the distribution of $R_{\rm mis}$ with a Rayleigh distribution:
\beqa\label{eq:mis_sig}
P(R_{\rm mis}) = \frac{R_{\rm mis}}{\sigma^2} \exp{\frac{R^2_{\rm mis}}{2\sigma^2}},
\eeqa
where for the variance of the distribution, $\sigma$, we assume that the miscentering distance is proportional to the halo virial radius: $\sigma = c_{\rm mis} R_{\rm vir}$. We will treat $\ln(c_{\rm mis})$ as a free parameter.  This expression captures the idea that larger halos can have larger miscentering distances, although we do not expect this form to hold exactly over all halo masses.

The miscentered profile after averaging across the probability distribution of $R_{\rm mis}$ is then:
\beqa
\xi^{\rm{halo}-y}_{\rm mis}(R) = \int dR_{\rm mis}P(R_{\rm mis})\xi^{\rm{halo}-y}(R|R_{\rm mis}).
\eeqa
The final miscentered profile is then given (on average) by a weighted sum of the correctly centered profile and the miscentered profile:
\beqa\label{eq:mis_f}
\xi^{\rm{halo}-y}_{\rm total}(R) = f_{\rm mis}\xi^{\rm{halo}-y}_{\rm mis}(R) + (1 - f_{\rm mis})\xi^{\rm{halo}-y}(R),
\eeqa
where $f_{\rm mis}$ is another free parameter that quantifies the fraction of miscentered halos. Lastly, we transform this miscentered profile back into harmonic space.

A similar formulation to the above has been used to describe the miscentering of optically identified galaxy clusters by \citet{Rykoff:2016} and others. Using tSZ and X-ray data, \citet{Rykoff:2016} found $\ln(c_{\rm mis}) = -1.13 \pm 0.22$ and $f_{\rm mis} = 0.22 \pm 0.11$.  As a way to roughly account for miscentering in our analysis, we assume that these constraints apply to all halos in our analysis.  This assumption effectively extrapolates the \citet{Rykoff:2016} results to low halo masses.  Given that our main intention is to qualitatively illustrate the impact of halo miscentering on our forecasts, and that the precise level of miscentering will depend on the detailed properties of a future halo sample, this approach is reasonable. The fiducial values and priors for the miscentering parameters $\ln(c_{\rm mis})$ and $f_{\rm mis}$ used in our analysis are given in Table \ref{tab:hy}. 

\subsubsection{Mass bias}

For the halo-based forecasts, we assume that the halo population can be divided into bins based on halo mass.  Of course, inferring halo masses is challenging, and may be subject to systematic errors; we refer to any difference between the true halo mass and the assumed halo mass as mass bias.  Perhaps the most powerful way to infer halo masses is to use gravitational weak lensing.  Other techniques, such as dynamical masses \citep[e.g][]{Farahi:2016} or masses inferred from clustering \citep[e.g.][]{Baxter:2016} can also be used.  With all of these techniques in mind, we assume somewhat conservatively that the halo masses can be calibrated at the level of 10\% precision.    This simple assumption suffices for our purposes, since our primary goal is to illustrate the level at which systematics will be important for future analyses, rather than constructing the most realistic forecast possible.  For comparison, the weak lensing analysis of \citet{McClintock:2019} calibrated the masses of cluster-scale dark matter halos with statistical precision of roughly 5\%, and a systematic error budget of roughly 5\%.  At the low-mass end, \citet{Lin:2016} constrained the mean mass of groups from the \citet{Yang:2007} catalog with $M \sim 10^{12} M_{\odot}/h$ to roughly 50--60\% precision. For the galaxy-based forecasts, rather than assuming a prior on halo masses, we will assume that galaxy clustering is used to constrain the HOD.

We assume that some observable quantity (for example, stellar mass) is used to put the halos into mass bins.  The mass distribution within the bin is then assumed to be described by a lognormal distribution:
\beq
\label{eq:massdistribution}
\frac{dP(M_{\rm obs} | M_{\rm true})}{d{\ln M}_{\rm obs}} = \frac{1}{\sqrt{2\pi \sigma^2_{\ln M}}} \exp{[-x^2]}
\eeq
where
\beq\label{eq:MB_x}
x = \frac{\ln(\eta M_{\rm obs}) - \ln(M_{\rm true})}{\sqrt{2 \sigma^2_{\ln M}}},
\eeq
where we fix $\sigma_{\rm lnM} = 0.5$, as a reasonable level of scatter in the mass-observable relation \citep[e.g.][]{Yang:2007}.  The parameter $\eta$ then controls the level of bias in this relation.  We impose a 10\% prior on $\eta$, as described above.  We then marginalize over the probability distribution in Eq.~\ref{eq:massdistribution} when computing the integrals over mass in Eqs.~\ref{eq:Cl1h}, \ref{eq:by}, \ref{eq:ng}, \ref{eq:bgl}, \ref{eq:bhl}, and \ref{eq:cov_trispec}.  

\subsubsection{Biases in the $y$ maps}

Another potential source of systematic error for the halo-$y$ correlation measurements is contamination of the Compton-$y$ maps by other sources of mm-wave emission, such as the cosmic infrared background (CIB) and radio point sources \citep{Planck:tsz}. Of these, the CIB is potentially the most problematic, as shown in e.g. \citet{Pandey:2019}, since it cannot be masked and tends to produce a positive response in component-separated Compton-$y$ maps~\cite{Planck:tsz,Hill-Spergel2014}.  In this analysis, we will make the simplifying assumption that bias from the CIB can be controlled to a level below the statistical errors.  For example, one can leverage high-frequency data to construct Compton-$y$ maps with CIB-like SEDs projected out~\cite{Remazeilles2011,Pandey:2019,Madhavacheril2019}, or in a stacking analysis one can simultaneously fit for the tSZ and infrared emission signals~\cite{Greco:2015}.

Moreover, this assumption is reasonable for most of our analysis because we focus on cross-correlations with low-redshift galaxies and halos with $z < 0.5$.  Since the CIB is sourced primarily at higher redshifts $z \sim 2$~\cite{Planck:cib}, it should not introduce large biases for these correlations.  However, when we consider halo-$y$ correlations at high redshift ($z \sim 1$) and the $y$ auto-correlation, possible CIB contamination is more of an issue.  

Still, given the currently large uncertainties in CIB modeling, we postpone a more detailed consideration of these biases to future work. We note that a robust method to deproject CIB from Compton-$y$ maps would allow one to study feedback processes at high redshift. In particular, Compton-$y$ cross-correlations with high-redshift quasars from current and future surveys can directly shed light on AGN feedback, but interpretations of these cross-correlations can be highly sensitive to CIB modelling (see, for example, \citet{Soergel2017} and references therein).

\subsection{Survey assumptions}
\label{sec:survey_assumptions}

\begin{figure}
    \centering
    \includegraphics[width=0.475\textwidth]{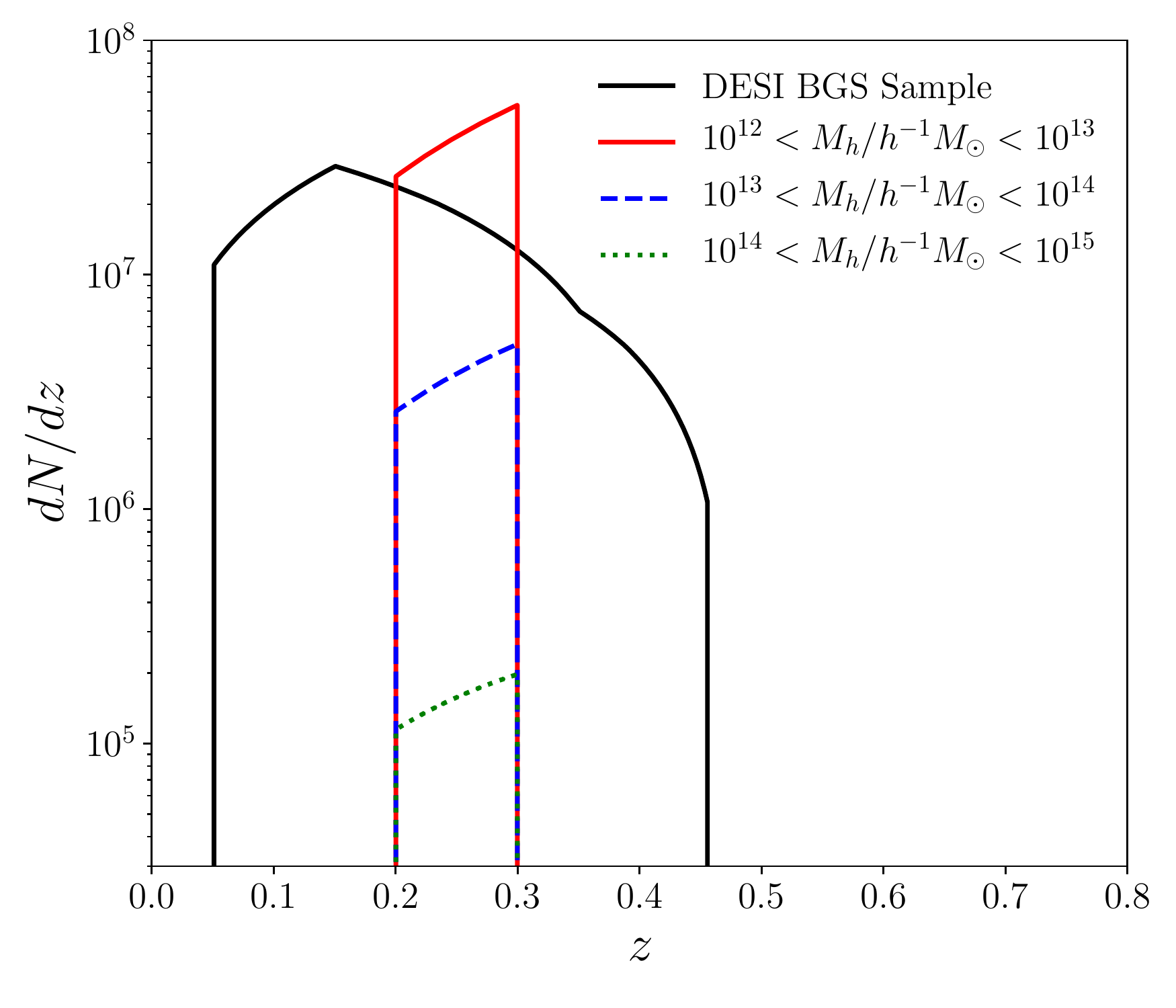}
    \caption{Redshift distributions of the halo and galaxy samples used for the forecasts. The colored lines indicate the redshift distributions of the halo samples in the three mass bins, and for the lowest redshift bin considered in this analysis ($0.2 < z < 0.3$).  The black curve shows the redshift distribution of the BGS galaxy sample from DESI used for the galaxy-based forecasts.  
    }
    \label{fig:desi_dndz}
\end{figure}

Measuring a galaxy-$y$ correlation requires both a galaxy survey and a CMB survey with which to construct maps of Compton-$y$.  We describe our survey assumptions in more detail below.

\subsubsection{CMB surveys}
\label{sec:cmb_surveys}

We focus on the future CMB-S4 survey~\cite{CMBS4DSR} and also present some results for the imminent Simons Observatory (SO) survey~\cite{Ade:2019} in this work, although ongoing ground-based CMB surveys (e.g., Advanced ACT~\cite{Henderson:2016} and SPT-3G~\cite{Benson:2014}) should also produce high-precision tSZ cross-correlation measurements in the near term.  The specifications of the SO and CMB-S4 surveys used here are described in Refs.~\cite{Ade:2019} and~\cite{CMBS4DSR}, respectively, to which we refer the reader for further details.  In brief, the frequency coverage of the SO and CMB-S4 large aperture telescope (LATs) is planned to be the same, including channels centered at 27, 39, 93, 145, 225, and 280 GHz.  In addition, the resolution of the SO and CMB-S4 LATs is expected to be the same, utilizing diffraction-limited optics on telescopes with a 6-meter primary dish (SO includes one such LAT, while CMB-S4 includes two LATs for the ``wide-field survey'' that is relevant to our study).  This yields a beam with FWHM = 1.4 arcmin at 145 GHz.  Full details of the noise modeling --- including both instrumental noise and non-white atmospheric noise with realistic frequency dependence --- for SO and CMB-S4 can be found in Refs.~\cite{Ade:2019} and~\cite{CMBS4DSR}, respectively.  As a rough guide, the expected white noise level of the SO LAT survey is 10 $\mu$K-arcmin (``baseline'' expectation)  at 145 GHz, while that of the CMB-S4 LAT wide-field survey is 2 $\mu$K-arcmin.  The SO LAT survey will cover 40\% of the sky; the CMB-S4 wide-field survey will cover 70\% of the sky, but we assume an effective area of 40\% for high-fidelity $y$-map reconstruction (our forecasts may thus be considered conservative in this sense).

In order to forecast realistic noise curves on the reconstructed Compton-$y$ parameter, we use the methodology described in Ref.~\cite{Ade:2019} for SO (see their Sec. 2) and in Ref.~\cite{CMBS4DSR} for CMB-S4 (see their Appendix A.3).  We note that Planck data from 30-353 GHz is also assumed to be included in the $y$ reconstruction, which is useful on large angular scales where atmospheric noise is significant for SO and CMB-S4.  This methodology includes realistic modeling of all major components of the mm-wave sky at every frequency under consideration, combined with the SO and CMB-S4 noise modeling described above (and white noise assumed for Planck).  These components are then propagated through a harmonic-space internal linear combination (ILC)~\citep[e.g.][]{Eriksen2004} method to obtain post-component separation noise curves for the Compton-$y$ map, $N_{\ell}^{yy}$, including the effect of residual foregrounds and noise due to the instrument and atmosphere.  The derived noise curves for SO, which is scheduled to begin collecting science data in 2022, are publicly available.\footnote{\url{https://simonsobservatory.org/assets/supplements/20180822\_SO_Noise\_Public.tgz}}  Due to possible evolution in the design of CMB-S4, which is scheduled to start in the late 2020s, we refer to the noise curves utilized here as ``CMB-S4-like''.  We utilize the range of angular scales from $80 < \ell < 8000$ in this analysis.

For simplicity, we use so-called ``standard'' ILC noise curves here, which simply minimize the total variance of all non-tSZ components in the final map (i.e., no component is required to vanish explicitly).  However, we note that future analyses may use ILC tSZ maps with CIB-like component SEDs projected out in order to mitigate possible biases.  This would modestly increase the noise in the derived $y$-map, and thereby increase the error bars on some of the forecasts presented here.  High-frequency data from e.g. CCAT-prime~\cite{CCAT} could be useful in mitigating these effects.  Given current uncertainties in CIB modeling, we defer a careful consideration of this issue to future work.

\subsubsection{Galaxy survey: halos}
\label{sec:halo_sample}

For the galaxy survey, we consider two types of forecasts.  The first assumes that the galaxy survey is used to identify an underlying population of dark matter halos.  The second type assumes that the galaxy-$y$ correlation is measured directly, and a HOD is used to relate the galaxies to the underlying halos that they populate. 

For the halo-based forecasts, we assume that halos have been identified down to some minimum halo mass, $M_{\rm min}$ and out to some redshift, $z_{\rm max}$.  Such catalogs have previously been constructed using spectroscopic survey by \citep[e.g.][]{Yang:2007}.  We assume that the completeness fraction of the halo catalog is unity for the mass and redshift bins that we analyze, i.e., within these bins and over the assumed sky coverage, {\it all} dark matter halos have been identified.  Any incompleteness would necessarily increase the error bars in our forecasts. Of course, a real survey is unlikely to have a completeness fraction equal to unity, but we make this choice for simplicity and since it makes the results easy to interpret.  For comparison, the group catalog constructed in \citet{Yang:2007} had an approximately constant completeness fraction greater than $\sim 85\%$ for halos with $M_h > 10^{12} M_{\odot}/h$ and $ z < 0.2$.  Our assumption of perfect completeness should therefore have a small impact on the accuracy of our forecasts.

We consider three halo mass bins : $[10^{12}, 10^{13}]$, $[10^{13}, 10^{14}]$ and $[10^{14}, 10^{15}]\, M_{\odot}/h$, and two halo redshift bins: $[0.2,0.3]$ and $[0.9, 1.2]$, although we focus on the lower redshift bin. The redshift distribution of halos in the lower redshift bin is shown in Fig.~\ref{fig:desi_dndz}. These bins are intended to very roughly reflect the mass and redshift coverage of the DESI survey.  The BGS sample of DESI peaks at around the redshift range of 0.2--0.3, and the emission line galaxy (ELG) sample peaks in the range 0.9--1.2 \citep{DESI:FDR}. 

Our results for the halo-based forecasts are sufficiently general that they can also be applied to different galaxy surveys, such as LSST \cite{LSST:sciencebook}.  For cluster-scale halo masses, LSST should enable the halo-$y$ cross-correlations to be measured over a broad redshift range ($z \lesssim 1$) and over a large area overlapping with CMB-S4 ($f_{\rm sky} \sim 0.3$).  Since LSST is a photometric survey, the construction of a low mass group catalog is perhaps more challenging and uncertain than for a spectroscopic survey.  Photometric cluster finders like redMaPPer \citep{redmapper} typically work best at high masses, $M \gtrsim 10^{14}\,M_{\odot}$.  Assuming LSST can be used to identify low-mass groups, it should enable high signal-to-noise measurements of halo-$y$ correlations.  Even without identifying groups, though, cross-correlations of LSST with Compton-$y$ maps should enable tight constraints on feedback models using the HOD framework that we discuss below.  We note that since $y$ is a line of sight projected quantity, modeling the halo-$y$ correlation function is not particularly sensitive to photometric redshift errors.  

For the halo sample we assume $f_{\rm sky} = 0.3$ when computing the covariance of the halo-$y$ cross-correlations.  This is a somewhat optimistic estimate of the overlapping area between DESI and SO/CMB-S4 surveys, and a slightly pessimistic estimate for the overlap of LSST and SO/CMB-S4.

\subsubsection{Galaxy survey: galaxies}
\label{sec:galaxy_sample}

Our galaxy-based forecast is designed to represent the BGS sample of DESI \citep{DESI:FDR}. We adopt the HOD model described in \S\ref{sec:hod_halo}. We choose the fiducial values of the HOD parameters to be equal to the best fit values for sub-sample of SDSS redshift survey galaxies having absolute magnitude less than -19.5 \citep{Zehavi:2011} since BGS galaxies are expected to have a similar maximum absolute magnitude \citep{DESI:FDR}.  The redshift distribution of this galaxy sample is shown in Fig.~\ref{fig:desi_dndz}.  When computing the covariance of the galaxy-galaxy auto-correlation and galaxy-$y$ cross-correlation, we assume $f_{\rm sky} = 0.23$, the estimated sky fraction for overlap of DESI with SO and CMB-S4 \citep{Ade:2019}.

We note that our forecasts are based on using the full halo and galaxy catalogs to perform the cross-correlation measurements. Some previous analyses \citep[e.g.,][]{Planck:2013:hy} have imposed isolation criteria on galaxy catalogs to reduce the contributions from the two-halo term.  We do not take this approach here, as we build a full model for the one and two-halo terms in the galaxy-$y$ correlation.  Moreover, since we are most interested in low-mass halos for which the two-halo term makes significant contributions to the halo-$y$ correlation at small  scales, following the isolation approach would require precisely modeling the impact of the isolation criterion.  Such modeling will be dependent on the exact isolation criterion imposed, which adds significant complications to the analysis~\citep{Hill:2018}.  Finally, by imposing an  isolation  criterion,  one can significantly reduce the number of halos used in the analysis, and thereby degrade signal-to-noise and the derived parameter constraints.

\begin{figure*}
    \centering
    \includegraphics[scale=0.6]{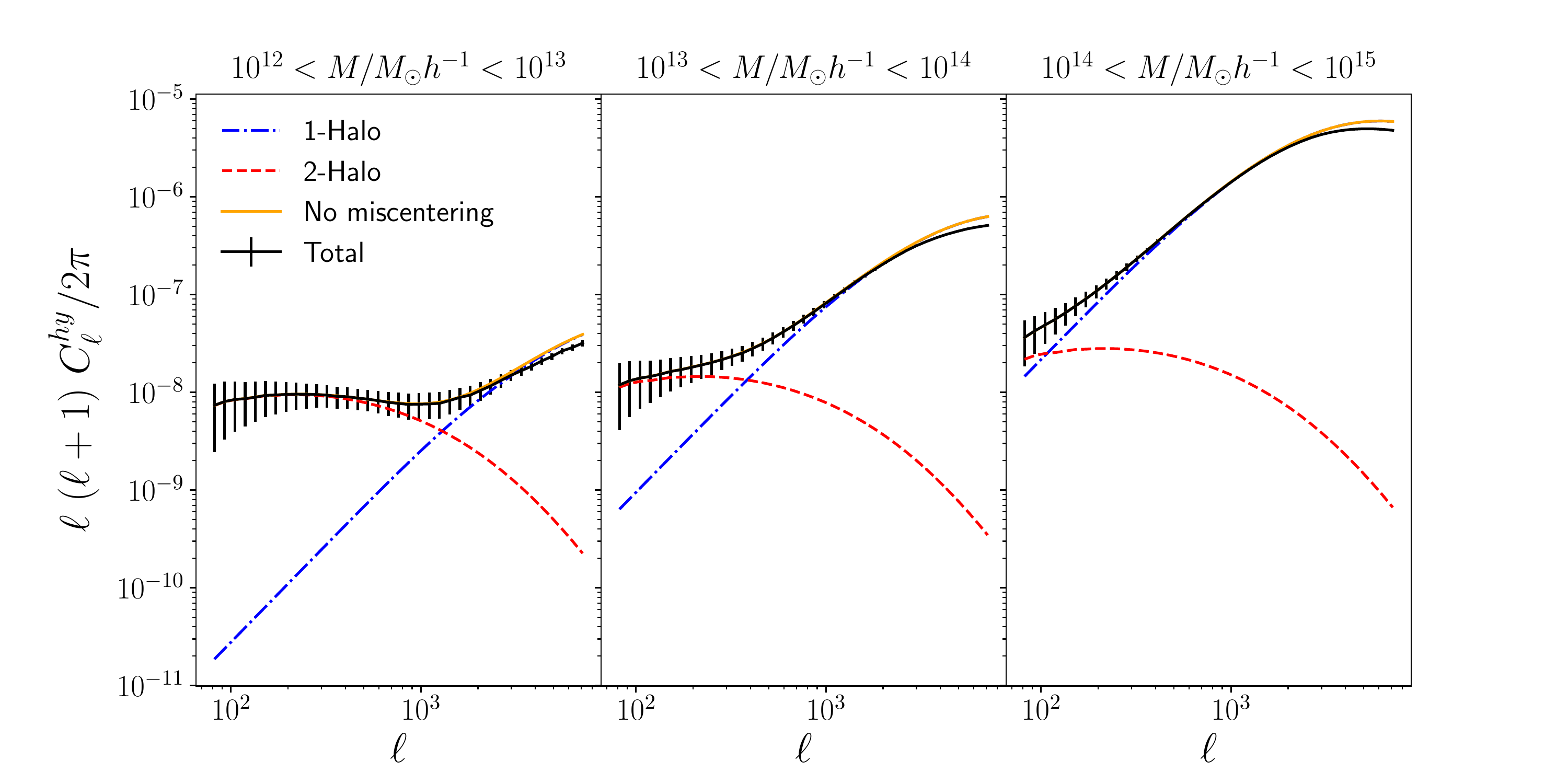}
    \caption{The halo-$y$ correlation functions and forecasted errorbars  (black) for halos with redshifts $0.2 < z < 0.3$.  We show the one-halo and two-halo contributions to the correlation functions with the colored curves.  The orange curve shows the model prediction in the absence of any miscentering.  Miscentering has the impact of reducing the cross-power at small scales; the impact of miscentering is detected at more than $50\sigma$ significance for the highest mass bin.  
    }
    \label{fig:Cells}
\end{figure*}

\subsection{Projected constraints}
\label{sec:likelihood}

To forecast future constraints we use a parameter fitting approach.  We adopt a Gaussian likelihood:
\begin{eqnarray}
\ln \mathcal{L}(\vec{d}|\theta) = -\frac{1}{2}(\vec{d} - \vec{m}(\theta))^T \varmathbb{C}^{-1} (\vec{d} - \vec{m}(\theta)),
\end{eqnarray}
where $\vec{d}$ is the vector of $C_{\ell}$ measurements, $\varmathbb{C}$ is their covariance, and $\vec{m}(\theta)$ is the model evaluated at parameter values, $\theta$.  The free parameters in the halo and galaxy-based forecasts are summarized in Tables~\ref{tab:hy} and \ref{tab:gy}, respectively.

The posterior on model parameters is then 
\begin{eqnarray}
\mathcal{P}(\theta | \vec{d}) = \mathcal{L}(\vec{d}|\theta) {\rm Pr}(\theta),
\end{eqnarray}
where ${\rm Pr}(\theta)$ are the priors on model parameters. We will consider several choices of priors below.  For the parameters describing the pressure profiles, we typically adopt non-informative priors.  In some cases, however, when a parameter is very weakly constrained by a particular observable, we will adopt informative top hat priors, as we discuss more below.  For the systematics parameters describing mass bias and miscentering, we adopt priors assuming that these parameters are constrained by other measurements, as described in \S\ref{sec:sys_all}.  The parameters and priors in the analyses of halo-$y$ and galaxy-$y$ correlations are summarized in Tables~\ref{tab:hy} and \ref{tab:gy}, respectively.  We generate (weighted) samples from the posterior using \texttt{Multinest} sampling algorithm \citep{Multinest} as implemented in \texttt{Cosmosis} \citep{Cosmosis} package.

\begin{figure*}
    \centering
    \includegraphics[scale=0.5]{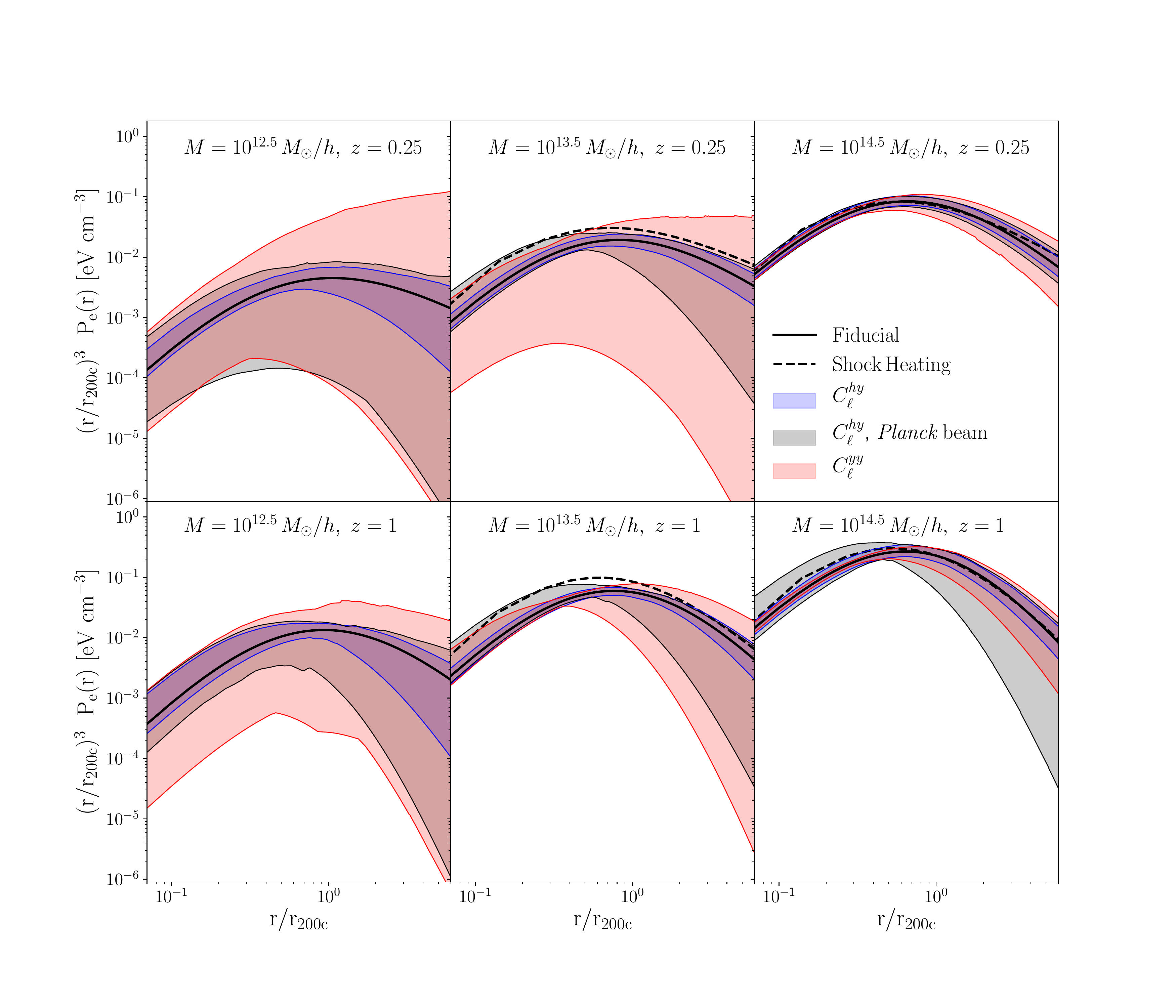}
    \caption{Forecasted constraints (2$\sigma$ uncertainty band) on the halo pressure profiles as functions of distance from the halo center.  The constraints are obtained from halos restricted to the mass bins $M_h \in [10^{12}, 10^{13}]$, $[10^{13}, 10^{14}]$ and $[10^{14}, 10^{15}]\, M_{\odot}/h$ (left to right), and two redshift bins $z \in [0.2,0.3]$, $[0.9,1.2]$ (top and bottom).  Given the measurements in these halo mass and redshift bins, the profiles are then evaluated at the indicated halo masses and redshifts.  The blue bands show the constraints recovered from the halo-$y$ cross correlation: these yield tight constraints on the halo pressure profiles for all masses and redshifts. The red band shows the constraints achieved with the Compton-$y$ autospectrum, which are significantly weaker than those from the halo-$y$ correlation at low halo mass and large $r$.  We also show (grey band) the constraints from the halo-$y$ correlation for a hypothetical experiment with the same noise level as CMB-S4, but with a 10 arcminute beam (i.e. the beam size of the {\it Planck} $y$-maps).  This band demonstrates that a small beam is essential to probing the pressure profiles of low-mass halos.  The solid line shows the fiducial pressure profile model, while the dashed line shows the pressure profile for the shock heating model from \citet{Battaglia:2010}, which does not include feedback, radiative cooling, or star formation. 
    }
    \label{fig:pressure_profile}
\end{figure*}

\section{Results I: halo-based forecasts}
\label{sec:halo_results}

We first consider forecasts for cross-correlations between the halo samples defined in \S\ref{sec:survey_assumptions} and future Compton $y$ maps.  Our main goal in this section is to illustrate several important aspects of pressure profile constraints derived from halo-$y$ correlations, including the impact of systematics.  In \S\ref{sec:galaxy_results} we will present forecasts for correlations between galaxies and Compton-$y$, marginalizing over the halo-galaxy connection.

\subsection{Forecasted signal-to-noise of halo-$y$ correlations}

Fig.~\ref{fig:Cells}, shows the model halo-$y$  spectra, including the one and two-halo components, for three different mass bins and for the redshift bin $0.2 < z < 0.3$.  Also shown are the forecasted errorbars for a CMB-S4-like experiment.  The projected total signal-to-noise for each of the mass bins is high, roughly $40\sigma$ for halos in the bin $[10^{12},10^{13}] M_{\odot}/h$, $210\sigma$ for $[10^{13},10^{14}] M_{\odot}/h$, and $510\sigma$ for $[10^{14},10^{15}] M_{\odot}/h$.  

The one halo term is detected at high significance, even for the mass bin with $[10^{12},10^{13}] M_{\odot}/h$.  For comparison, the \citet{Vikram2017} measurements of correlation between SDSS groups with {\it Planck} $y$-maps  detected the one-halo term down to a minimum mass of roughly $10^{13} M_{\odot}$.  The improvement in the CMB-S4 forecasts relative to the {\it Planck} measurements is driven by two factors: improvement in the Compton-$y$ map signal-to-noise, and decrease in the beam size, from 10 arcmin for {\it Planck} (set by the resolution of its 100 GHz channel) to roughly 1--2 arcmin for CMB-S4.  As we show below, the improvement in beam size is essential for high significance detections of the one-halo term at low halo mass.

We also forecast significant detection of the two-halo term: roughly $14\sigma$ for halos in the bin $[10^{12},10^{13}] M_{\odot}/h$, $10\sigma$ for $[10^{13},10^{14}] M_{\odot}/h$, and $6\sigma$ for halos with $[10^{14}, 10^{15}] M_{\odot}/h$ (all for the redshift bin of $0.2 < z < 0.3$).  In contrast to the one-halo term, the two-halo term is best detected around low-mass halos because its amplitude scales weakly with halo mass, and because the less massive halos are significantly more numerous. For comparison, using galaxy catalogs from year one data from DES and {\it Planck} $y$-maps, \citet{Pandey:2019} detected the two-halo term at roughly 3 to $5\sigma$ for several redshift bins.  The two halo term is an interesting-observable, as it is sensitive to the halo-bias-weighted pressure of the Universe, and can be used to probe the total thermal energy in halos at a given redshift \citep{Vikram2017, Pandey:2019}.

Fig.~\ref{fig:Cells} also shows the impact of halo miscentering on the Compton-$y$ correlations (see the difference between the orange and black curves).  Miscentering results in a suppression of power at small scales, and a smaller increase in power at larger scales (not visible in the plot given the large range of the $y$-axis).  Because the small scale measurements of the halo-$y$ correlation have very high signal-to-noise, if unaccounted for, miscentering would lead to a highly significant bias in parameter constraints.  For instance, for the $[10^{14},10^{15}]M_{\odot}/h$ mass bin, the impact of miscentering can be detected at roughly $50\sigma$.  We discuss degeneracy between the miscentering and pressure profile parameters in more detail below.

\begin{figure*}
    \centering
    \includegraphics[scale=0.7]{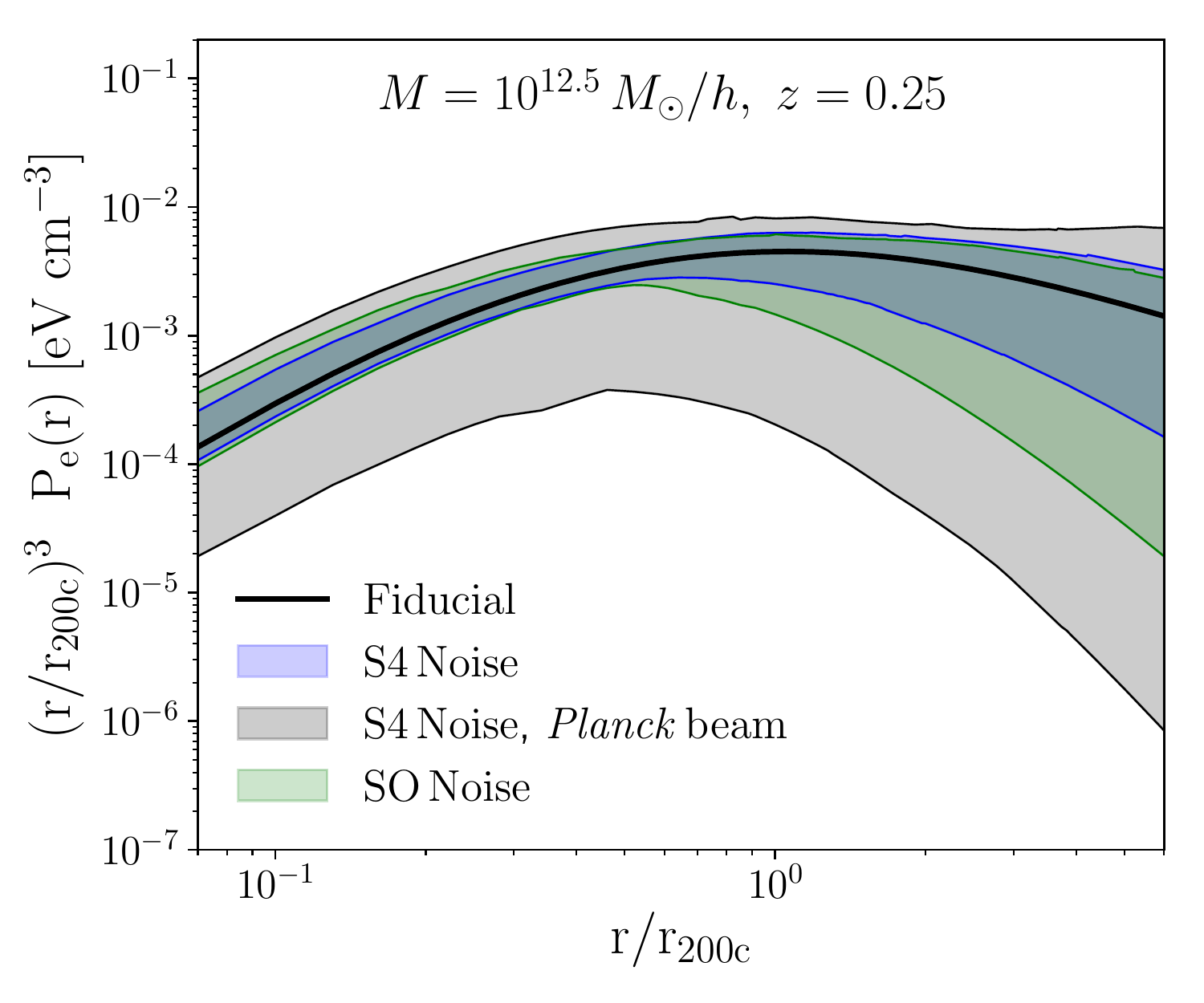}
    \caption{Forecasted constraints (2$\sigma$ uncertainty band) on the halo pressure profiles of low-mass halos from SO (green) and CMB-S4 (blue).  Also shown are constraints from a hypothetical experiment with the same noise level as CMB-S4, but with a {\it Planck}-like beam.  A small beam is essential to constraining the pressure profiles of low-mass halos.}
    \label{fig:pressure_profile_wSO}
\end{figure*}

\begin{figure*}
    \centering
    \includegraphics[scale=0.5]{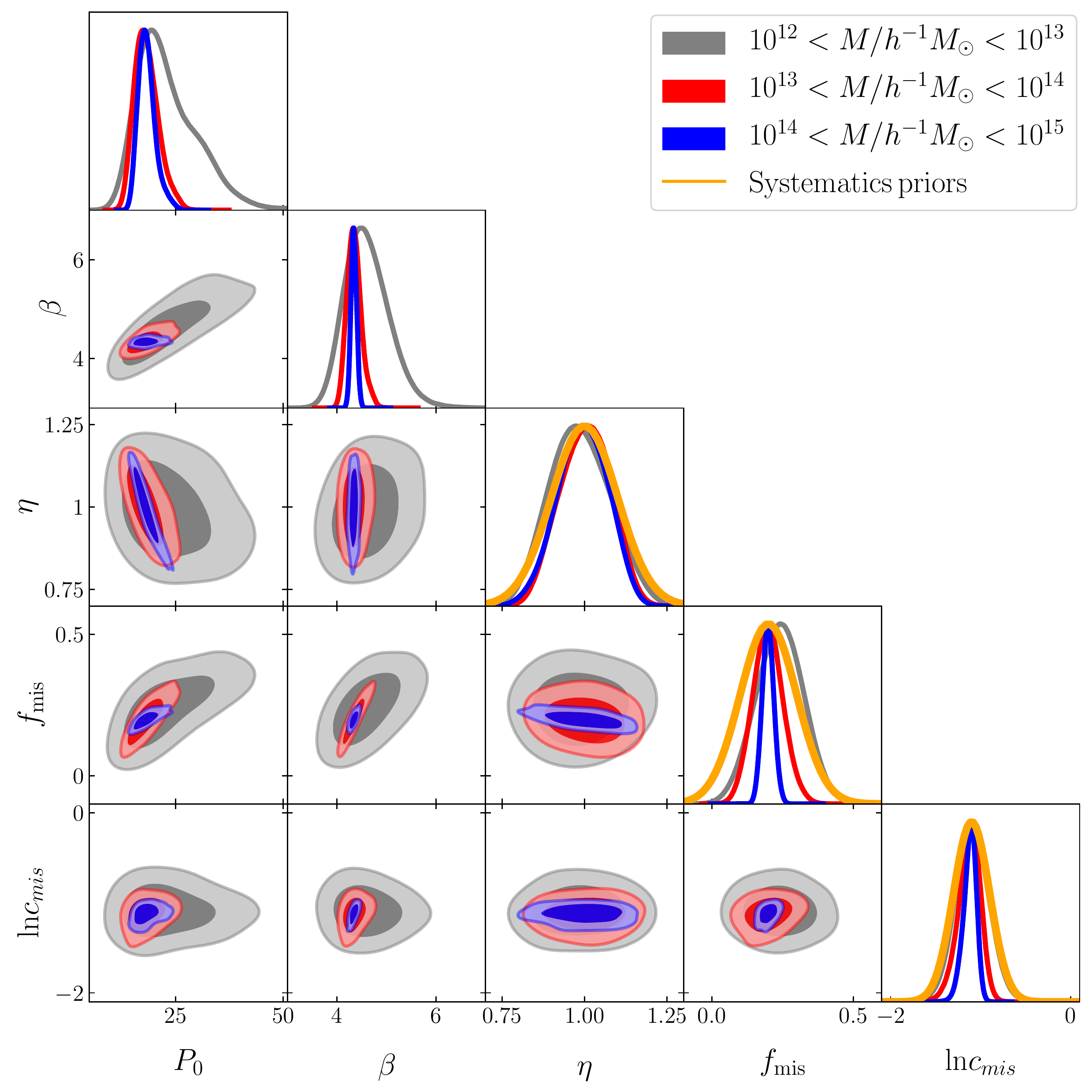}
    \caption{Constraints on the pressure profile and systematics parameters (Table \ref{tab:hy}) obtained from analyzing the halo-$y$ correlation in various halo mass bins.  The orange curves for $\eta$, $f_{\rm mis}$ and $\ln c_{\rm mis}$ indicate the priors on these parameters.  For high mass halos, the pressure profile amplitude ($P_0$) and shape ($\beta$) can be measured robustly, breaking degeneracy between these two parameters.  Additionally, for this mass bin, the miscentering parameters can be constrained using the data alone (i.e. without resorting to an informative prior).  For all halo mass bins, the mass bias parameter is prior dominated.  For high-mass halos, however, the prior on $\eta$ limits the degree to which the pressure profile amplitude can be constrained.  For low mass halos, the one-halo term is detected at lower signal-to-noise, and consequently, $P_0$ and $\beta$ are more degenerate.  As a result, improving the halo mass calibration beyond the assumed 10\% prior will not yield much better pressure profile constraints for these halos.  For low mass halos, priors on the miscentering fraction are more important than mass calibration.}
    \label{fig:halo_contours}
\end{figure*}

\subsection{Projected constraints on pressure profiles}\label{sec:pressure_constraints}

We now use the model fitting formalism discussed in \S\ref{sec:formalism} to forecast how well future measurements will constrain model parameters and the inferred 3D pressure profiles.  We note that a model fitting approach is essential to (a) using the 2D Compton-$y$ measurements to make inferences about the 3D pressure profiles, and to (b) account for the impact of systematics such as miscentering and mass bias.  We first consider forecasts at fixed cosmology; we will open up the cosmological parameter space in \S\ref{sec:cosmology_dependence}.

\begin{figure*}
    \centering
    \includegraphics[scale=0.7]{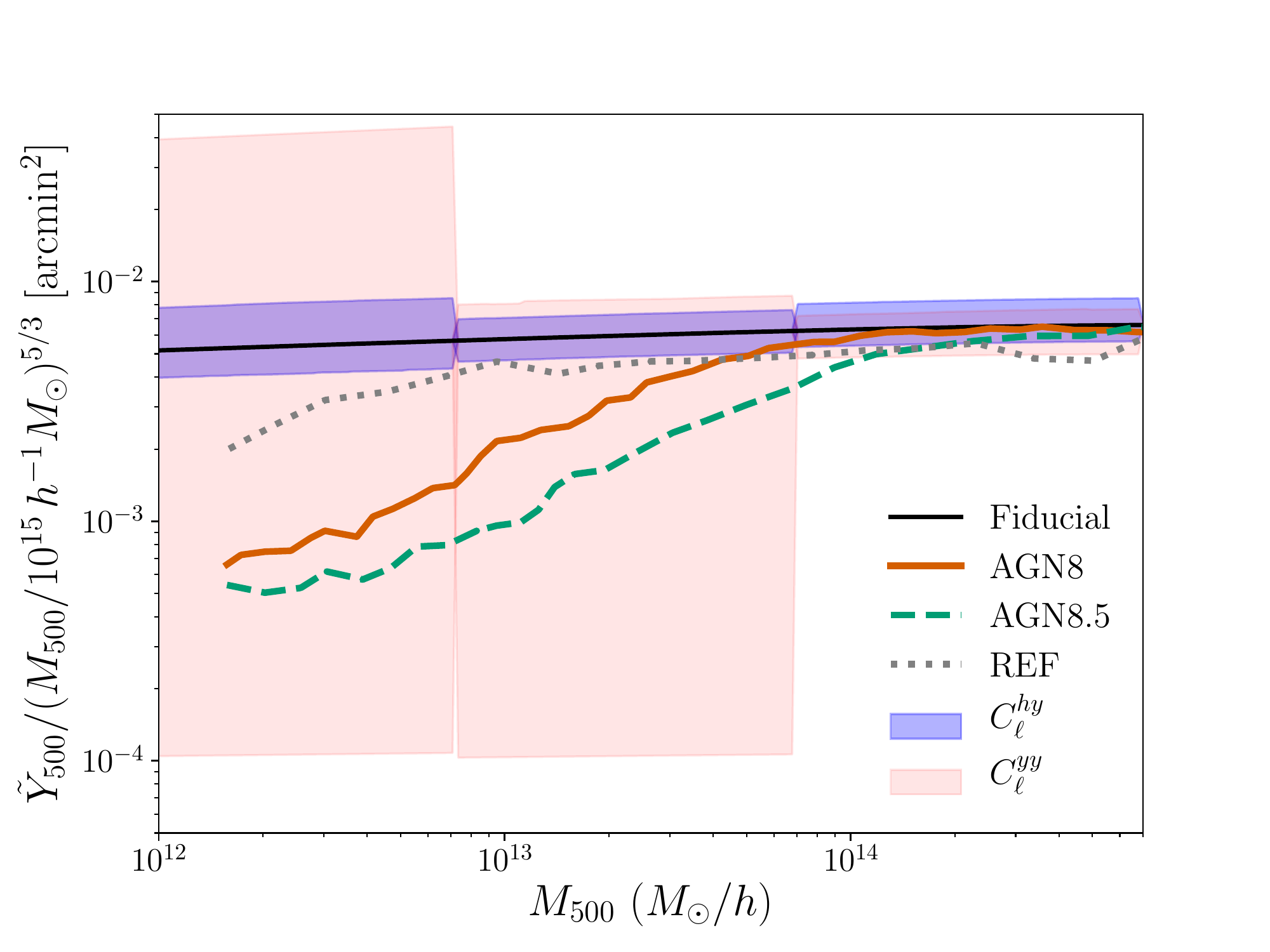}
    \caption{Forecasted constraints (2$\sigma$ uncertainty) on the $\tilde{Y}$-$M$ relation (see Eq.~\ref{eq:y500}) for the halo-$y$ cross-correlation and for the $y$ autospectrum.  Constraints are shown at $z=0.25$.  The assumed survey properties are described in \S\ref{sec:survey_assumptions} and correspond to a DESI-like galaxy survey, and a CMB-S4-like CMB survey; we assume that the halos are in a redshift bin with $0.2 < z < 0.3$. At low halo mass, the halo-$y$ correlation leads to significantly tighter constraints on the $\tilde{Y}$-$M$ relation than the $y$ autospectrum.  The constraints from the $y$ autospectrum are weak enough that the priors imposed on the pressure profile parameters are somewhat informative; this is not the case for the halo-$y$ correlations. The various colored curves show the $\tilde{Y}$-$M$ relation predicted from the hydrodynamical simulations analyzed in \citet{LeBrun15}, each with a different feedback prescription, see text for more details. }
    \label{fig:ym}
\end{figure*}

Fig.~\ref{fig:pressure_profile} shows the  constraints on the 3D pressure profiles for three halo mass bins and two redshift bins, as inferred from the forecasted halo-$y$ correlation measurements.   We have generated these forecasts while varying the parameters shown in Table~\ref{tab:hy} (with the priors described therein) for each halo mass bin. In all cases, tight constraints on the halo pressure profiles are achieved, despite marginalizing over the miscentering and mass bias models (see blue bands).  

For comparison, we also show in Fig.~\ref{fig:pressure_profile} the constraints obtained on the halo pressure profiles from the Compton-$y$ autospectrum (red bands).  
In order to make a fair comparison between the contraints from $C_{\ell}^{hy}$ and $C_{\ell}^{yy}$, we vary the $P_0$ and $\beta$ parameters only over the redshift bin of the halo sample when making projections for $C_{\ell}^{yy}$.  Because the Compton-$y$ autospectrum is dominated by the contributions from the most massive halos (see, e.g., Fig. 3 of \cite{Makiya:2018}), the $y$ autospectrum measurements are unable to constrain the pressure profiles of low mass halos.  We note that because the $y$ autospectrum constraints are very weak at low mass, these constraints are impacted to some degree by our choice of parameter priors.  We also note that the Compton-$y$ autospectrum does not depend on halo miscentering or halo mass bias, since no halo catalog is necessary for these measurements. 

Also shown in Fig.~\ref{fig:pressure_profile} is the pressure profile for the shock heating model from \citet{Battaglia:2010}.  This model includes no prescription for feedback or radiative cooling, so the difference between this curve and the the fiducial model provides some measure of the impact of these effects on the halo pressure profile.  For the halo mass bin $[10^{13},10^{14}] M_{\odot}/h$, the measurements at small radius can probe feedback at high signal-to-noise.  We do not show the shock heating curves for the lowest mass bin since this would require significant extrapolation of the \citet{Battaglia:2010} results.

Fig.~\ref{fig:pressure_profile_wSO} compares the pressure profile constraints on low-mass halos that can be obtained from SO to those projected for CMB-S4.  Both experiments provide similar constraints on the pressure profiles of low mass halos, given their substantial improvement in beam size and map depth over {\it Planck}.  The similarity of the SO and CMB-S4 constraints at low halo mass is due to degeneracy between $\beta$ and $P_0$, as we discuss below.  At high halo mass, the constraints from SO and CMB-S4 again end up similar because of degeneracy between $P_0$ and $\eta$.  Given the similarity of the SO and CMB-S4 forecasts, for simplicity, we will focus on CMB-S4 below. 

Fig.~\ref{fig:halo_contours} illustrates the degeneracies between the model parameters for several different halo mass bins.  The orange curves in Fig.~\ref{fig:halo_contours} illustrate the priors on the miscentering and mass bias parameters.  Several points are worth emphasizing about the degeneracies between the model parameters.  First, we note that at high halo mass, the shape ($\beta$) and amplitude ($P_0$) parameters of the pressure profiles are not very degenerate.  This is because at high masses, the one-halo term is robustly detected out to large radii.  For the low halo masses, on the other hand, there is less information about the halo outskirts, making it easier to trade off changes in the profile shape with changes to the profile amplitude, resulting in significant $P_0$-$\beta$ degeneracy.

We next consider the constraints on the pressure profile amplitude and its degeneracy with the mass bias parameter, $\eta$. For all halo mass bins, the constraints on mass bias are prior dominated, with the prior at the level of 10\% mass calibration (see discussion in \S\ref{sec:survey_assumptions}).  This prior is sufficient to obtain useful constraints on pressure profiles, as seen for instance in Fig.~\ref{fig:pressure_profile} and below.  As discussed above, a 10\% prior on halo masses is obtainable with current and future weak lensing surveys.  At low halo mass, significant improvements to the prior on $\eta$ would not significantly improve the pressure profile constraints (i.e. the posterior on $P_0$ and $\beta$ would remain roughly the same).  This is because of the large degeneracy between $P_0$ and $\beta$ for the low halo masses.  Given this large degeneracy, a 10\% prior on halo masses is essentially good enough for exploiting the pressure profile information in the data.

We next consider degeneracy between the pressure profile parameters and the miscentering parameters.  The miscentering fraction, $f_{\rm mis}$, is significantly degenerate with $P_0$ and $\beta$ for all halo masses.  This is not surprising, since as seen in Fig.~\ref{fig:Cells}, miscentering reduces the amplitude of the one-halo term, similar to the impact of changing $P_0$ or $\beta$.  Interestingly, though, the constraints on the miscentering parameters are not prior dominated for the high mass bins; in other words, the halo-$y$ correlations are sufficient to self-calibrate miscentering at high mass.  As discussed in \S\ref{sec:miscentering}, the miscentering model that we have adopted in this analysis is very approximate at low halo mass.  Still, it is clear that in order to obtain tight constraints on the pressure profiles of low-mass halos, one must understand the degree to which the assumed halo centers reflect the underlying centers, and that achieving this understanding may require data beyond the halo-$y$ correlation measurements themselves.

\begin{figure*}
    \centering
    \includegraphics[scale=0.5]{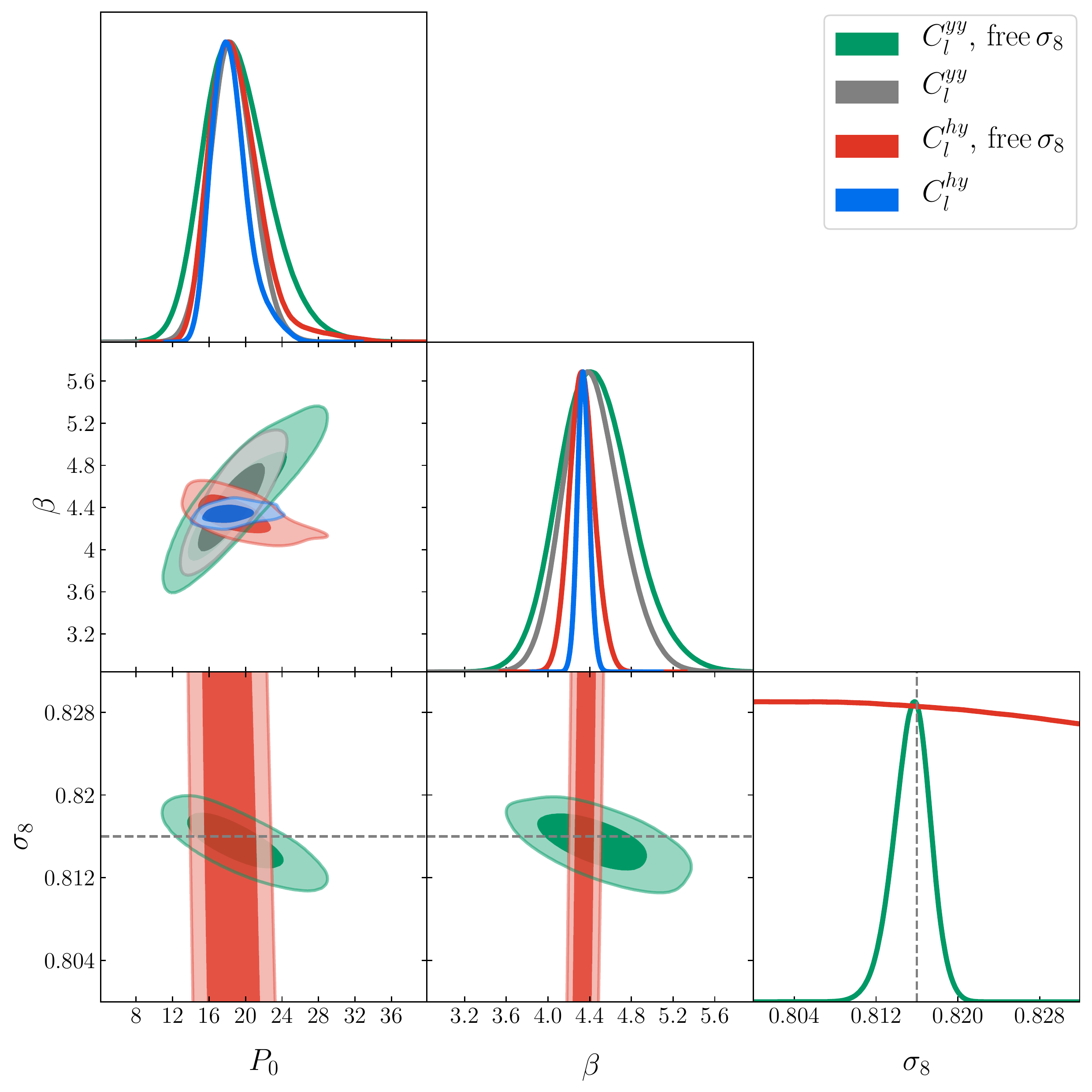}
    \caption{The impact of allowing freedom in $\sigma_8$ on the pressure profile constraints from the halo-$y$ cross-correlation measurements.  The constraints on pressure profile parameters from the $y$ autospectrum are very degenerate with $\sigma_8$; those from the halo-$y$ cross-correlation, on the other hand, are only weakly degenerate with $\sigma_8$.  The dashed grey line in the bottom panels indicates the fiducial value of $\sigma_8$.}
    \label{fig:halo_contours_cosmology}
\end{figure*}

\begin{figure*}
    \centering
    \includegraphics[scale=0.5]{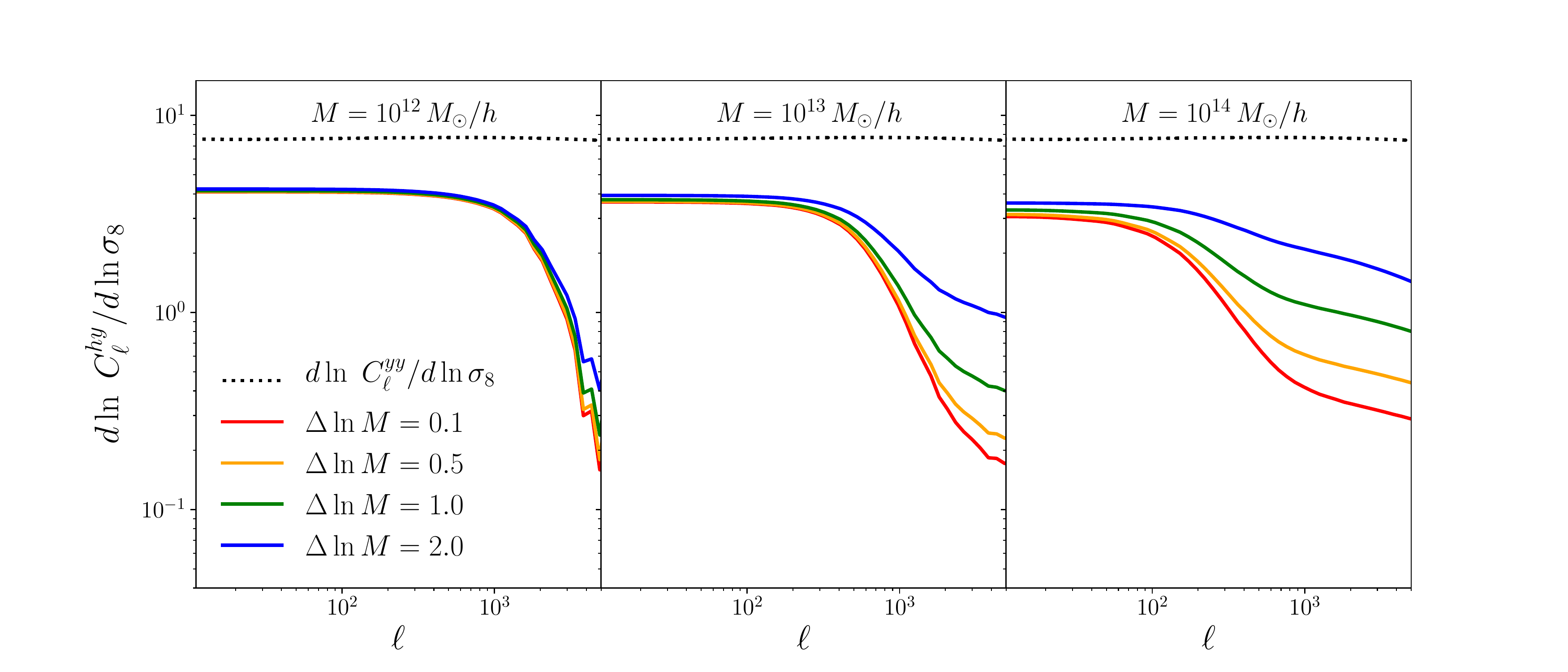}
    \caption{The dependence of $C^{hy}_{\ell}$ (solid curves) and $C_{\ell}^{yy}$ (dotted curve) on $\sigma_8$ for different values of logarithmic bin width ($\Delta \ln M$) at different central halo masses indicated in the panels. The high-mass halos are significantly one-halo dominated at large $\ell$, and hence sensitivity to $\sigma_8$ increases with increasing bin width (see text for more explanation). Interestingly, low-mass halos are mostly two-halo term dominated and hence the dependence of the sensitivity to $\sigma_8$ on the bin width is suppressed. For comparison, the sensitivity of $C^{yy}_{\ell}$ is depicted as a dotted line, which is roughly constant for all $\ell$. 
    }
    \label{fig:dCldsig8}
\end{figure*}

\subsection{Constraints on $\tilde{Y}$-$M$ relation}\label{sec:YM}

In addition to the pressure profiles of halos, it is also interesting to consider how well future surveys will constrain the integrated Compton-$y$ parameter as a function of halo mass.  In Fig.~\ref{fig:ym} we show forecasted constraints on the $\tilde{Y}$-$M$ relation for several mass bins, where $\tilde{Y} = \tilde{Y}_{500}$ is defined as

\beq\label{eq:y500}
\tilde{Y}_{500}(M,z) = \frac{D^2_A(z) }{(500 {\rm Mpc})^2 E^{2/3}(z)} 
\frac{\sigma_T}{m_e c^2} \int_0^{r_{500}} dr 4\pi r^2 \frac{P_e(r|M,z)}{D^2_A(z)}.
\eeq
Here, $D_A(z)$ is angular diameter distance out to redshift $z$, $E(z)$ is the dimensionless Hubble parameter, and $r_{500}$ is the radius of halo that encloses a mass having mean density of 500 times the critical density at redshift $z$. Note that the constraints shown for the halo-$y$ correlation correspond to treating each mass bin {\it independently}; constraints from a joint analysis of all mass bins simultaneously would necessarily be tighter. 

Fig.~\ref{fig:ym} additionally shows the forecasted constraints on the $\tilde{Y}$-$M$ relation from the analysis of the $y$ autospectrum (red bands).  The $y$ autospectrum is mostly sensitive to halos with $M > 10^{13}\,M_{\odot}$, with some dependence on redshift and $\ell$.   Consequently, at low mass, the $y$ autospectrum cannot constrain the $\tilde{Y}$-$M$ relation.  At high halo mass, however, the $y$ power spectrum yields tight constraints on the $\tilde{Y}$-$M$ relation. 

Also shown in Fig.~\ref{fig:ym} are predictions for the $\tilde{Y}$-$M$ relation from the analysis of cosmo-OverWhelmingly Large Simulation (cosmo-OWLS) suite of cosmological hydrodynamical simulations \citep{LeBrun:2014, McCarthy:2014} for different feedback models as described in \citet{LeBrun15}. The $\bf{REF}$ model incorporates the prescriptions of radiative cooling and supernovae feedback while $\bf{AGN8}$ and $\bf{AGN8.5}$ additionally include the feedback from AGN growth. The $\bf{AGN8.5}$ simulation results in more violent and episodic feedback mechanisms compared to $\bf{AGN8}$. The signal-to-noise of the $\tilde{Y}$-$M$ constraints from the halo-$y$ correlations is sufficient to distinguish between these models at high significance.  Future measurements of these correlations that probe low-mass halos will provide a powerful test of current hydrodynamical simulations and theoretical models.  We note that the impact of these different feedback models on the matter power spectrum varies by about 10\% at $k = 5 \,h/{\rm Mpc}$ (see e.g. Fig.~1 of \citet{Huang:2019}).  

While Fig.~\ref{fig:ym} presents the impact of feedback on the $\tilde{Y}$-$M$ relation as a function of halo mass, one could also consider the impact of feedback as a function of halo redshift.  Because the halos can be restricted to narrow redshift slices, $C_{\ell}^{hy}$ provides a potentially powerful handle on the redshift evolution of feedback.  Moreover, as shown in Fig.~\ref{fig:pressure_profile}, the constraints on the pressure profiles of high-redshift halos are not much worse than those at low redshift.  We postpone a detailed exploration of this possibility to future work.

\subsection{Cosmology dependence}
\label{sec:cosmology_dependence}

We now consider the impact of allowing freedom in both the cosmological model and the pressure profile model.  For the purposes of illustration, we consider in this section only varying $\sigma_8$, to which the $y$-autospectrum and $y$ cross-correlation observables are very sensitive~\cite[e.g][]{Komatsu:2002,Hill-Pajer2013,Bolliet2019}.

Fig.~\ref{fig:halo_contours_cosmology} shows the impact of allowing freedom in $\sigma_8$ when fitting the halo-$y$ and $y$-$y$ spectra.  The $y$ autospectrum is extremely sensitive to $\sigma_8$, leading to a strong degeneracy between the pressure profile parameters and $\sigma_8$.  The halo-$y$ correlations, on the other hand, are less sensitive to $\sigma_8$, as can be seen from the red contours in the bottom panels of Fig.~\ref{fig:halo_contours_cosmology}.  

The insensitivity of the pressure profile constraints from $C_{\ell}^{hy}$ to $\sigma_8$ can be understood as follows.  In the limit of an infinitely narrow mass bin, $C_{\ell}^{hy}$ in the one-halo regime is simply the pressure profile of the halos.  The pressure profile of halos is not expected to be sensitive to the amplitude of the matter density fluctuations.  However, since $C_{\ell}^{yy}$ is an integrated quantity over halos of all the masses, it will be extremely sensitive to $\sigma_8$ through the halo mass function (Eq.\ref{eq:Cl1h} and \ref{eq:Cl2h}). This is the behavior we see in Fig.~\ref{fig:halo_contours_cosmology}. 

It is also interesting to consider how the sensitivity of $C_{\ell}^{hy}$ to $\sigma_8$ depends on the width of the halo mass bin.  This is shown in Fig.~\ref{fig:dCldsig8}. At high mass and for a narrow bin, at high $\ell$ where the one-halo contribution to $C^{hy}_{\ell}$ dominates, the cross-spectrum is simply the profile of the halos, and hence the sensitivity to $\sigma_8$ is low. But, as is clear from the plot, the one-halo sensitivity of $C_{\ell}^{hy}$ to $\sigma_8$ increases with the width of the bin, since the bin averages over halos of different mass and is therefore sensitive to the halo abundance.  In the limit that the bin is infinitely wide, one would imagine that the sensitivity of $C_{\ell}^{hy}$ to $\sigma_8$ should approach that of $\sqrt{C_{\ell}^{yy}}$ (since it depends on one less power of $y$).  This behavior can be seen in Fig.~\ref{fig:dCldsig8}, where the dependence of the $y$-autospectrum on $\sigma_8$ is roughly $\sigma_8^8$, and the dependence of the halo-$y$ cross-spectrum approaches $\sigma_8^4$ in the limit of a wide mass bin.  In the two-halo regime, the dependence on $\sigma_8$ is only weakly impacted by the changing width of the mass bin, since the halo bias depends weakly on the bin width.  

\section{Results II: galaxy-based forecasts}
\label{sec:galaxy_results}

We now consider forecasts for cross-correlations of galaxies with maps of Compton y.  The forecasts in this section adopt the galaxy sample model described in \S\ref{sec:galaxy_sample}.  The projected signal-to-noise of these correlations is very high: roughly 108$\sigma$.  

Unlike the halo based forecasts described in \S\ref{sec:halo_results}, the results described in this section are subject to the additional complexity of relating the galaxy sample to the underlying halos that they populate.  Because of this uncertainty, we find --- not surprisingly --- that the galaxy-$y$ correlation alone is insufficient to yield tight constraints on the halo pressure profiles.  Consequently, in this section we consider parameter constraints for the joint analysis of $C_{\ell}^{gy}$ and $C_{\ell}^{gg}$.

\begin{figure*}
    \centering
    \includegraphics[scale=0.7]{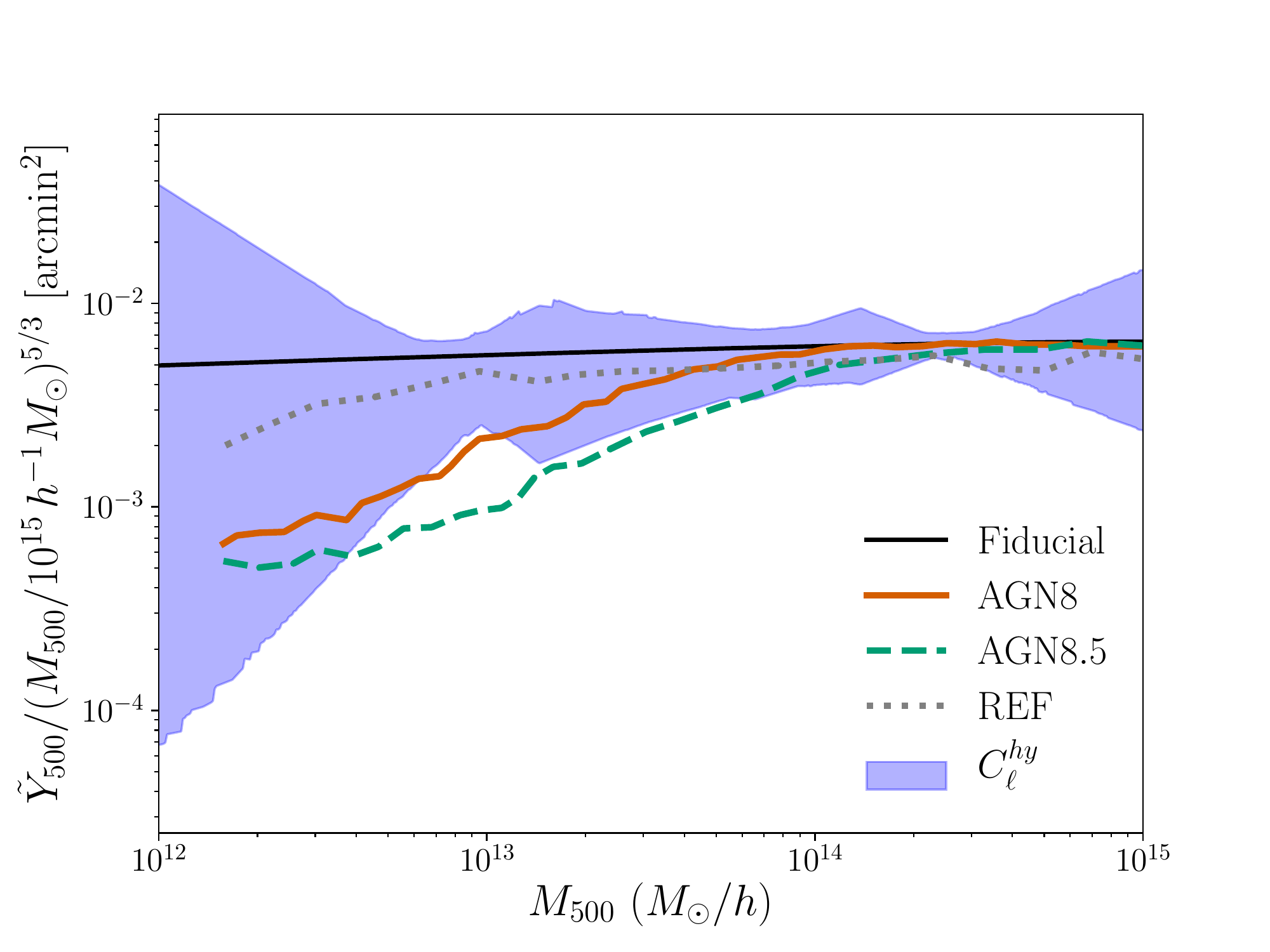}
    \caption{The 2-$\sigma$ uncertainty band on the $\tilde{Y}$-$M$ relation, as inferred from cross-correlations between DESI BGS galaxies and Compton-$y$ maps from CMB S4 (using $C^{gg}_{\ell}$ and $C^{gy}_{\ell}$). The curves show the $\tilde{Y}$-$M$ relation predicted from  hydrodynamical simulations, as presented in \citet{LeBrun15}. 
    }
    \label{fig:desi_YM}
\end{figure*}

As discussed in \S\ref{sec:pressure}, when modeling the galaxy-$y$ correlation, we adopt a variation on the \citetalias{Battaglia:2012} fitting formulae that introduces additional freedom into the model at low mass. Table \ref{tab:gy} provides a summary of all the parameters varied in this case and the choice of prior ranges. Using this model and generating posterior samples as discussed in \S\ref{sec:likelihood}, we show forecasts for the constraints on the halo $\tilde{Y}$-$M$ relation in Fig.~\ref{fig:desi_YM}.  These constraints are marginalized over the HOD parameters describing the galaxy sample.  Despite this marginalization, galaxy-$y$  correlations with a DESI-like survey will yield tight constraints on the $\tilde{Y}$-$M$ relation down to low halo mass.  The resultant sensitivity will be sufficient to distinguish between different AGN feedback models, especially when combining constraints across a range of masses. 

\begin{figure*}
    \centering
    \includegraphics[scale=0.5]{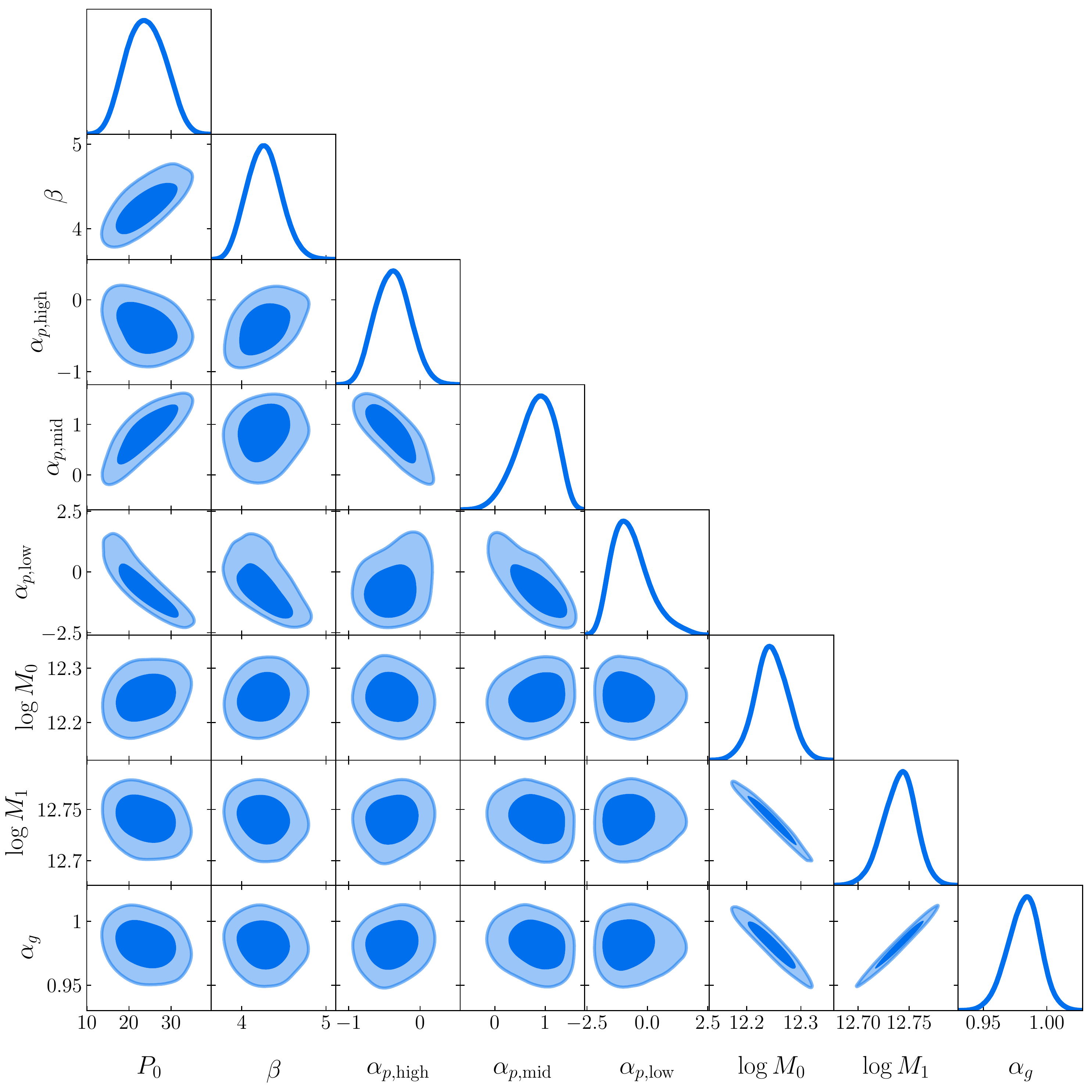}
    \caption{Constraints on model parameters from joint fits to the galaxy-$y$ and galaxy-galaxy correlations.  The meanings of the parameters as well as the choice of priors are described in Table~\ref{tab:gy}. We see that the parameters related to the pressure profile are not significantly degenerate with the HOD parameters.  This lack of degeneracy is because the clustering measurements ($C_{\ell}^{gg}$) tightly constrain the HOD parameters.}
    \label{fig:galaxy_contours}
\end{figure*}

We show constraints in the full parameter space in Fig.~\ref{fig:galaxy_contours}, where we find that there is little degeneracy between the pressure profile parameters and the HOD parameters.  This is because the HOD constraints from $C_{\ell}^{gg}$ are tight enough that the residual uncertainty on the HOD parameters does not significantly degrade the pressure profile constraints.  This is encouraging: we can determine halo pressure profiles accurately in the future without worrying that they will be degraded by HOD uncertainty. 

We note that there are two significant differences in the $\tilde{Y}$-$M$ constraints obtained from the halo and galaxy-based forecasts.  First, the halo forecasts treat the halos in each mass bin separately; the galaxy-based forecasts, on the other hand, consider contributions to the signal from all halo masses simultaneously (see discussion in \S\ref{sec:pressure}).  Second, because the halo forecasts consider each mass bin separately, we only free an amplitude parameter ($P_0$) and a shape parameter ($\beta$) for each bin.  For the galaxy forecasts, since we fit a large range of halo masses simultaneously, we allow must introduce freedom into the $\tilde{Y}$-$M$ via the $\alpha_p$ parameters in Eq.~\ref{eq:B12_model}, \ref{eq:B12_scaling} and \ref{eq:Pe_total}.  For these reasons, one should be careful in comparing the $\tilde{Y}$-$M$ constraints in Fig.~\ref{fig:ym} to those of Fig.~\ref{fig:desi_YM}.

\section{Discussion}
\label{sec:discussion}

We have made forecasts for future measurements of cross-correlations between dark matter halos (and the galaxies that populate them) and maps of the Compton-$y$ parameter.  Future galaxy surveys (e.g. DESI and LSST) and CMB surveys (e.g. SO and CMB-S4) will enable measurements of these correlations at very high signal-to-noise, upwards of $500\sigma$ in some cases.  Particularly exciting is that next generation CMB observations will be higher resolution than current {\it Planck} measurements, enabling the pressure profiles of low-mass halos to be probed.  We forecast signal-to-noise of roughly $50\sigma$ for halos in the mass range of $[10^{12}, 10^{13}] M_{\odot}/h$, and show that tight constraints on the baryonic pressure profiles of these halos can be achieved out to high redshift, even in the presence of systematics like miscentering and mass bias (Fig.~\ref{fig:pressure_profile}).  Cross-correlations between galaxies and halos are essential to achieving interesting constraints on the pressure profiles: by itself, the Compton-$y$ autospectrum yields poor constraints on the pressure profiles of low-mass halos (see e.g. Fig.~\ref{fig:pressure_profile}).

Even after including important systematic effects like halo mass bias and miscentering, we find that tight constraints on halo pressure profiles can be achieved.  These constraints will be sufficient to distinguish between different models of baryonic feedback (see Fig.~\ref{fig:ym} and Fig.~\ref{fig:desi_YM}).  This is an exciting prospect for several reasons.  For one, understanding feedback is essential for a complete understanding of galaxy formation.  Additionally, uncertainty on feedback models currently limits the ability of weak lensing surveys to exploit signal-to-noise available at small physical scales.  Tighter constraints on feedback models from galaxy-$y$ correlations should therefore lead to tighter constraints on cosmology. 

We note that feedback can also be probed with observations of thermal X-rays emitted by the hot gas in halos.  However, the tSZ effect offers several unique advantages over X-ray observations.  For one, the tSZ effect is approximately redshift-independent (while the X-ray surface brightness declines as $(1+z)^4$), enabling tSZ constraints on feedback at high redshift.  The tSZ signal also scales more slowly with halo mass than the X-ray signal, enabling tSZ observations to probe lower-mass halos, for which feedback  is expected to be particularly important. 

The high signal-to-noise forecasted in this work for the halo-$y$ correlations also means that one can split the galaxy samples to study how feedback processes correlate with galaxy type.  For example, at fixed halo mass, one might expect red galaxies to exhibit stronger signs of feedback (e.g., gas blown out to large radii or gas at higher temperatures) than blue galaxies, as feedback processes are thought to be responsible for injecting energy and quenching star formation (thereby driving the transition from blue to red colors)~\cite{Spacek2017}.  Beyond galaxies, we also note that cross-correlations of quasar samples with Compton-$y$ maps could provide direct evidence of ongoing feedback activity, as has been sought for in studies with {\it Planck}~\cite{Verdier2016,Soergel2017} and ACT data~\cite{Crichton2016,Hall2019}.

While we have focused on projections for future surveys in this analysis, measurements of halo-$y$ correlations and constraints on pressure profile parameters have already been achieved from correlations of e.g. SDSS and {\it Planck} \citep{Vikram2017, Hill:2018} and DES and {\it Planck} \citep{Pandey:2019}. Upcoming analyses that cross-correlate DES with Compton-$y$ maps from ACT \citep{Swetz2011} and SPT \citep{Carlstrom:2011} will extend the reach of these measurements to lower halo mass.  Additionally, while we have focused on the thermal SZ effect in this analysis, the kinematic SZ (kSZ) effect can also be used to probe feedback.  By combining the kSZ and tSZ measurements around halos, one can constrain the level of non-thermal pressure support in the halos \citep{Battaglia:2017, Battaglia:2019}, which can be significantly impacted by feedback processes. 

\acknowledgements

We thank Bhuvnesh Jain, Mike Jarvis, Adam Lidz, and an anonymous referee for useful discussions and feedback on an early draft of this work.  JCH acknowledges support from the W. M. Keck Fund at the Institute for Advanced Study.  This is not an official Simons Observatory Collaboration paper. SP is supported by the U.S. National Science Foundation through award AST-1440226 for the ACT project; both EJB and SP are partially supported by the US Department of Energy grant DE-SC0007901.

\bibliographystyle{mnras} 
\bibliography{ref}

\end{document}